\documentclass[11pt,english]{article}
\usepackage[T1]{fontenc}
\usepackage[latin9]{inputenc}
\usepackage{float}
\usepackage{amsmath}
\usepackage{amssymb}
\usepackage{graphicx}
\usepackage{geometry}
\usepackage{setspace}
\usepackage{esint}
\makeatletter

\providecommand{\tabularnewline}{\\}
\floatstyle{ruled}
\newfloat{algorithm}{tbp}{loa}
\providecommand{\algorithmname}{Algorithm}
\floatname{algorithm}{\protect\algorithmname}

\newtheorem{theorem}{Theorem}

\newtheorem{condition}{Condition}

\newtheorem{lemma}{Lemma}

\newenvironment{proof}[1][Proof]{\textbf{#1.} }{\ \rule{0.5em}{0.5em}}

\makeatother

\usepackage{babel}
\usepackage{comment}
\usepackage{caption}
\usepackage{subfig,float}
\usepackage[labelfont={bf,sf}]{caption}
\usepackage{graphicx}

\makeatother

\usepackage{babel}
\begin{document}
\title{Estimation of an Order Book Dependent Hawkes Process for Large Datasets\thanks{We are very grateful to the Editor Dacheng Xiu and the Referees for
their comments that have led to substantial improvements both in content
and presentation. We are also grateful to Francesco Cordoni (Royal
Holloway), Giuliano De Rossi (Goldman Sachs) and Yuri Taranenko (ADIA)
for useful conversations related to the topic of this paper. }}
\author{Luca Mucciante\thanks{Department of Economics, Royal Holloway University of London, Egham
TW20 0EX, UK. Email: lucahost@gmail.com.} \and Alessio Sancetta\thanks{Corresponding author. Department of Economics, Royal Holloway University
of London, Egham TW20 0EX, UK. Email: asancetta@gmail.com.}}
\date{July 20, 2023}
\maketitle
\begin{abstract}
A point process for event arrivals in high frequency trading is presented.
The intensity is the product of a Hawkes process and high dimensional
functions of covariates derived from the order book. Conditions for
stationarity of the process are stated. An algorithm is presented
to estimate the model even in the presence of billions of data points,
possibly mapping covariates into a high dimensional space. The large
sample size can be common for high frequency data applications using
multiple liquid instruments. Convergence of the algorithm is shown,
consistency results under weak conditions is established, and a test
statistic to assess out of sample performance of different model specifications
is suggested. The methodology is applied to the study of four stocks
that trade on the New York Stock Exchange (NYSE). The out of sample
testing procedure suggests that capturing the nonlinearity of the
order book information adds value to the self exciting nature of high
frequency trading events. 


\end{abstract}

\section{Introduction}

This paper presents an intensity model for event arrivals in high
frequency trading. The intensity depends on order book information.
The model is a Hawkes process where the intensity does not only depend
on the time from an event arrival but also on the order book. The
model is specifically designed for estimation in the presence of high
dimensional covariates. Moreover, we present an estimation procedure
that can deal with large datasets in a relatively simple way, using
quadratic programming. High frequency data that include the order
book may contain more than a million records in a day for a liquid
instrument. When we consider information from other instruments, we
may easily have millions of updates per day. 

The problem of estimating the intensity of trading events conditional
on order book and trade data relies to a single realization of trade
and order book orders rather than a cross-section, as in the case
of hazard models. The case when the number of conditioning variables
is large has been studied recently (Sancetta, 2018, Mucciante and
Sancetta, 2022). However, these references ignore the self exciting
nature of event arrivals which is well documented by a number of authors
(Bacry et al., 2015, Filimonov and Sornette, 2015 for reviews). The
high dimensional model in Sancetta (2018) accounts for self excitement
in the intensity in a simulation study, but does not provide a proof
of the stationarity and ergodicity of the process. 

Recently, the statistical properties of Hawkes processes that incorporate
some information from the order book have been studied by a number
of authors (inter alia, Fosset et al., 2020, Morariu-Patrichi and
Pakkanen, 2022, Mounjid et al., 2019, Wu et al., 2019). These references
define a probabilistic model for some order book information. For
example they can be used to model the arrival of limit, market and
cancel orders conditioning on the queue size of the order book. These
models are general, but also rather complex when the dimension of
the conditioning events grows. The state variables tend to be restricted
to a finite set, which for practical reasons need to be low dimensional.
Hence, applications usually focus on say one order book variable taking
a finite number of discrete values. Hence, they are not suited for
estimation, conditioning on a large information set. Moreover, the
approach does not lend itself to the test of functional restriction
on the impact that the order book variables can have on the intensity. 

The focus of the current paper differs from the above in a number
of ways. First, we allow for the covariates that capture the effect
of the order book to take values in a subset of the real line rather
than taking values in a finite state space. Second, the number of
covariates can be large in the order of hundreds if not thousands.
This is important, as the number of levels in the order book can be
large. For examples, when considering the first ten levels of the
order book, we have ten values for prices and ten for quantities for
the bid and the ask, respectively. When we include information from
additional instruments and add covariates that capture the time dynamics,
it is easy to see how the information set can grow fast. Third, we
chose a parametrization that can model nonlinearities mapping covariates
into a higher dimensional space. This is particularly suited for high
dimensional estimation and testing restrictions. 

In summary, the main focus of the present paper is on modelling and
consistent estimation, allowing for possibly complex nonlinear impact
of order book variables and high dimensional information sets. The
goal is not to model the order book, but to use information from the
order book as possible predictors when modelling the intensity of
high frequency event arrivals. adds to the literature in a complementary
way. 

We now introduce the model. Let $N:=\left(N\left(t\right)\right)_{t\geq0}$
be the number of trade arrivals adapted to a filtration $\mathcal{F}=\left(\mathcal{F}_{t}\right)_{t\geq0}$.
The time of the $j^{th}$ event arrival is denoted by $T_{j}$, $j\geq1$
with $T_{0}:=0$. The counting process admits an $\mathcal{F}_{t}$-adapted
stochastic intensity $\lambda_{0}\left(t\right)$ such that

\begin{equation}
\lambda_{0}\left(t\right)=h_{0}\left(t\right)g_{0}\left(t\right),\label{EQ_intensityRepresentation}
\end{equation}
where $h_{0}$ is the predictable process 
\begin{equation}
h_{0}\left(t\right)=c_{0}+\int_{\left(-\infty,t\right)}\left(\sum_{l=1}^{L}d_{0,l}e^{-a_{0,l}\left(t-s\right)}\right)dN\left(s\right)=c_{0}+\sum_{j\geq0:T_{j}<t}\sum_{l=1}^{L}d_{0,l}e^{-a_{0,l}\left(t-T_{j}\right)}\label{EQ_hawkes_ht}
\end{equation}
with $a_{0,l},c_{0,l},d_{0,l}\geq0$, and
\begin{equation}
g_{0}\left(t\right)=X\left(t\right)'b_{0},\label{EQ_g0Environment}
\end{equation}
where $b_{0}$ is a positive $K\times1$ vector, $X:=\left(X(t)\right)_{t\geq0}$
is a positive $K\times1$ dimensional left-continuous process; throughout,
the prime symbol $'$ stands for transposition. The positivity condition
on $b_{0}$ and $X$ ensures that (\ref{EQ_intensityRepresentation})
is positive. The intensity can be interpreted as $\lambda_{0}\left(t\right)=\lim_{s\downarrow0}\Pr\left(N\left(t+s\right)-N\left(t\right)>0|\mathcal{F}_{t}\right)/s$
so that $M\left(t\right)=N\left(t\right)-\int_{-\infty}^{t}\lambda_{0}\left(s\right)ds$
is an $\mathcal{F}_{t}$-martingale. In this paper $K$ is assumed
to be large possibly in the order of thousands. Throughout, for identification
reasons, $d_{0,1}:=1$.

When (\ref{EQ_g0Environment}) is constant, (\ref{EQ_intensityRepresentation})
reduces to a Hawkes process, where the kernel is the sum of exponential
functions. It is well known that Hawkes processes provide a clear
interpretation in terms of exogenous information arrival, captured
by $c_{0}$, and endogenous market activity resulting from the second
term in (\ref{EQ_hawkes_ht}) (Hawkes and Oakes, 1974). However this
interpretation misses any information on the trading environment.
The latter is captured by the order book and other state variables
as summarized in (\ref{EQ_g0Environment}). We interpret (\ref{EQ_g0Environment})
as the market environment. Section \ref{Section_exogenousEndogenousTrading}
in the Appendix provides additional remarks on the model interpretation.
Our interest is on the impact of the state variables on the system.
This impact can be non-linear and we wish to capture these nonlinearities
in a way that is easy to interpret. To do so, in applications, $X$
may map a vector valued covariate process $Z:=\left(Z\left(t\right)\right)_{t\geq0}$
into a high dimensional space, where $Z$ represents information from
the order book. The empirical application of this paper focuses on
a special type of mapping that is called one-hot encoding and is popular
in machine learning due to its ease of interpretation (Alaya et al.,
2019). Other examples include series expansions such as Bernstein
polynomials, estimation in reproducing kernel Hilbert spaces etc.
One hot-encoding essentially discretizes the variables and maps them
into dummy variables. The cost of increasing the dimensionality of
the problem comes at the advantage of having variables that are easy
to interpret and to which linear constraints such as monotonicity
can be imposed in a natural way. For the model in (\ref{EQ_intensityRepresentation})
we state conditions for stationarity and ergodicity.

The model needs to be estimated under the constraints that we have
briefly outlined: very large datasets, high dimensional parameters,
and parameter's restrictions. We present an estimation procedure that
alternate between estimation of $h_{0}$ for fixed $g$ and uses surrogate
loss function for estimation of $g_{0}$ for fixed $h$. Mutatis mutandis,
our approach is close in spirit to estimation by coordinate descent,
which is standard in many problems (Friedman et al., 2007) and has
sound theoretical properties (Beck and Tetruashvili, 2013). We then
show that we can consistently estimate the high dimensional parameter
$b_{0}$ as long as $T^{-1}\ln K\rightarrow0$, where $\left[0,T\right]$
is the interval over which we collect the sample data. 

The empirical application is based on Level 3 data for four stocks
and the ETF on the S\&P500 as auxiliary instrument. All data are traded
on the NYSE. The data has been accessed through the Lobster dataset
(Huang and Polak, 2011). Level 3 data allows us to correctly synchronize
trades and order book. In the application, we estimate the model,
and are interested in establishing the shape of the non-linear impact
of variables derived from the order book. These include order book
volume imbalance and spread among others. We also include information
from other instruments. 

We observe that our non-negativity constraint on $b_{0}$ in (\ref{EQ_g0Environment})
leads to a form of ``shrinkage''. Most of the coefficients in the
estimation are zero due to the non-negativity constraint. This is
what we mean by ``shrinkage'', and it does not require to explicitly
employ a penalty on the coefficients. This phenomenon has been formally
studied in Mucciante and Sancetta (2022). We shall briefly comment
on this in Section \ref{Section_remarksRegularityConditions}.

\subsection{Further Remarks on the Literature\label{Section_additionalIntroRemarks}}

The use of intensity models in high frequency econometrics was pioneered
by Engle and Russell (1998). A review of different methodologies can
be found in Bauwens and Hautsch (2009). As a result of data availability,
there has been increasing interest in order book data on top of transaction
data. High frequency trading strategies rely on order book features.
For example, the literature has found that order book volume imbalances
and other order book variables have an impact on price movements and
trade arrivals at very short term horizons (Hall and Hautsch, 2007,
Cont et al., 2014, Sancetta, 2018). MacKenzie (2017) reports anonymous
interviews with ex algorithmic traders of the market maker Automated
Trading Desk. These interviews confirm the importance of order book
information for price movements. The empirical application confirms
the importance of this within the context of our model.

The estimation of the intensity for hazard models conditioning on
a large number of covariates has been considered by Ga\"{i}ffas and
Guilloux (2012). The main difference is that that the problem of estimating
the intensity of trading events conditional on order book and trades
information relies on a single realization of trade and order book
orders rather than a cross-section. 

A model similar to (\ref{EQ_intensityRepresentation}) with (\ref{EQ_hawkes_ht})
as baseline intensity has been considered in Sancetta (2018). However
that approach cannot easily scale to a number $m$ of covariate updates
in the order of millions or more and is not as easily interpretable
as the current method. When $h_{0}$ is constant, Sancetta and Mucciante
(2022) propose a consistent estimation method when the dimension $K$
of $b_{0}$ is large relatively to the number of jump events, with
no need to impose a penalty. However, they require the variables to
be linearly independent. This is not necessarily the case in many
large scale problems. Moreover, they do not allow for estimation of
a baseline intensity $h_{0}$. The method in that paper is complementary
to the current one and provides theoretical insights into the shrinkage
aspect of the procedure. 

There is a rich literature that deals with estimation using large
datasets, for example using distributed computing and averaging estimators
across different smaller subsamples (Zhang et al., 2015). However,
the structure of our problem is such that we can use all the data
in one single estimation, at least as far as estimation of $g_{0}$
in (\ref{EQ_g0Environment}) is concerned, and it benefits from good
asymptotic properties. Estimation of $h_{0}$ is a low dimensional
problem. Given $g_{0}$, estimation of $h_{0}$ is equivalent to estimation
of a Hawkes process. Despite being low dimensional, it may still pose
challenges as first recognized by Ogata and Akaike (1982), and recently
documented by Filimonov and Sornette (2015). Amongst other reasons,
alternative procedures to likelihood estimation of Hawkes processes
have been proposed in the literature (Da Fonseca and Zaatour, 2014,
Kirchner, 2017, Cartea et al., 2021). We do not attempt to address
these problems here. However, using our quadratic estimating function
for $h_{0}$, estimates appeared relatively stable.

\subsection{Outline of the Paper}

Section \ref{Section_Model} states the regularity conditions for
estimation of the model as well as conditions for stationarity and
ergodicity (Theorem \ref{Theorem_stationarityErgodicity}). Section
\ref{Section_estimation} discusses the estimation challenges and
presents an algorithm to address these. Section \ref{Section_simulationsConvergence}
contains a simulation to assess the validity of the estimation algorithm
and its finite sample performance. The focus is mostly on its convergence,
but as a byproduct, we also show that it possesses the variable screening
property under appropriate conditions. Asymptotic results concerning
consistency when $b_{0}$ is high dimensional are presented in Section
\ref{Section_asymptoticResults}. We extend the validity of the test
procedure discussed in Sancetta (2018) to compare the fit of two competing
intensities to the case of unbounded intensities. The model in (\ref{EQ_intensityRepresentation})
has unbounded intensity. The empirical application to four liquid
stocks traded on the NYSE can be found in Section \ref{Section_empirical}.
There, we show that accounting for order book information and its
dynamics adds value. Moreover, we find that the resulting impact of
covariates derived from the order book is nonlinear and can be remarkably
regular in some cases. Proofs and additional details can be found
in the Appendix. 

\section{The Model \label{Section_Model}}

Here, we introduce the regularity conditions used for estimation of
the model. We also consider a subset of these conditions to show that
a slight generalization of our model is a stationary and ergodic process.

\subsection{Regularity Conditions}

We use $\left|\cdot\right|_{1}$ to denote the $\ell_{1}$ norm of
a vector and $\left|\cdot\right|_{\infty}$ for the uniform norm $\left|g_{0}\right|_{\infty}=\sup_{t\geq0}\left|g_{0}\left(t\right)\right|$.
In order to deduce stationarity of the process, we need some form
of weak exogeneity of $X\left(t\right)$ conditioning on $\left(N_{s}\right)_{s<t}$.
We shall show that this is not restrictive. To introduce the exogeneity
condition, we need some additional notation. Let $\mathbb{M}$ be
the space of Radon measures on $\mathbb{R}$. Define $S_{t}$ to be
the operator on $\mathbb{M}$ that shifts mass $t$ units to the left:
$S_{t}\nu\left(\cdot\right)=\nu\left(t+\cdot\right)$ for $\nu\in\mathbb{M}$.
A map $f$ from $\mathbb{M}$ to the reals is causal if $f\left(\nu\right)=f\left(\tilde{\nu}\right)$
whenever $\nu=\tilde{\nu}$ on $\left(-\infty,0\right)$, $\nu,\tilde{\nu}\in\mathbb{M}$.
We write $\tilde{\nu}\succeq\nu$ if $\tilde{\nu}\left(C\right)\geq\nu\left(C\right)$
for any $C\subset\mathbb{R}$. Then, we call $f:\mathbb{M}\rightarrow\mathbb{R}$
nondecreasing if $f\left(\tilde{\nu}\right)\geq f\left(\nu\right)$
for any $\tilde{\nu}\succeq\nu$ such that $\nu,\nu\in\mathbb{M}$.
With slight abuse of notation, let $N\left(C\right)=\int_{C}dN\left(t\right)$
for any $C\subset\mathbb{R}$. Hence, here we view $N$ as an element
in $\mathbb{M}$: the random measure associated to a point process.
Let $\left(W\left(t\right)\right)_{t\geq0}$ be a predictable stationary
and ergodic process with values in $\mathbb{R}^{l}$, for some positive
integer $l$, such that $W\left(t\right)$ is independent of $N\left(t+C\right)$
for any $C\subseteq\left(-\infty,0\right)$. We refer to this as the
independence condition.

\begin{condition}\label{Condition_endogeneity} (Weak Exogeneity)
There is a strictly positive $f_{0}:\mathbb{R}^{l}\times\mathbb{M}\rightarrow\mathbb{R}$
such that for any $x\in\mathbb{R}^{l}$, $f_{0}\left(x,\cdot\right)$
is nondecreasing and causal and such that $g_{0}\left(t\right)=f_{0}\left(W\left(t\right),S_{t}N\right)$,
where the latent process $W$ is stationary and ergodic, and satisfies
the independence condition.\end{condition}

In the above, we do not need $g_{0}\left(t\right)=X\left(t\right)'b_{0}$.
We now specialize the conditions for the purpose of estimation. 

\begin{condition}\label{Condition_trueModel} (True Model) The intensity
of the process $N$ is not identically zero, is as in (\ref{EQ_intensityRepresentation}),
$X:=\left(X\left(t\right)\right)_{t\geq0}$ is left continuous and
takes values in $\left[0,1\right]^{K}$ for every $t\geq0$, $b_{0}$
has non-negative entries and is such that $\left|b_{0}\right|_{1}\leq B$,
where $B\sum_{l=1}^{L}\frac{d_{l}}{a_{l}}<1$ and $d_{0,1}:=1$.\end{condition}

\begin{condition}\label{Condition_g}(Set $\mathcal{G}$) $\mathcal{G}=\left\{ g=X'b:\left|b\right|_{1}\leq B,b_{k}\geq0,k=1,2,...,K\right\} $,
$g_{0}\in\mathcal{G}$.\end{condition}

\begin{condition}\label{Condition_h}(Set $\mathcal{H}$) We have
that $\mathcal{H}:=\left\{ h_{\psi}:\psi\in\Psi\right\} $ where 
\[
h_{\psi}\left(t\right)=c+\int_{\left(-\infty,t\right)}\left(\sum_{l=1}^{L}d_{l}e^{-a_{l}\left(t-s\right)}\right)dN\left(s\right),
\]
$\psi=\left\{ c,d,a\right\} $, $\Psi=\mathcal{C}\cup\mathcal{D}\cup\mathcal{A}$
where $\mathcal{C}\subset\left(0,\infty\right)$, $\mathcal{D}\subset\left[0,\infty\right)^{L}$
and $\mathcal{A}\subset\left(0,\infty\right)^{L}$ are compact sets
that contain $c_{0}$, $d_{0,l}$ and $a_{0,l}$ as in (\ref{EQ_hawkes_ht}).
\end{condition}

We refer to the above as the Regularity Conditions. Note that the
true model is in the sets used for estimation. Here, $B$ is a free
parameter that needs to be tuned. For reasons to be discussed in Section
\ref{Section_choiceB}, its choice is not crucial. In Theorem \ref{Theorem_stationarityErgodicity}
in Section \ref{Section_stabilityMultiplicativeHawkes}, we show that
$B\sum_{l=1}^{L}d_{0,l}/a_{0,l}<1$ together with the weak exogeneity
implies that the counting process is stationary. 

\subsection{Remarks on Regularity Conditions\label{Section_remarksRegularityConditions}}

We remark on regularity conditions. 

\paragraph{Condition \ref{Condition_endogeneity}.}

We show how Condition \ref{Condition_endogeneity} applies within
our framework, i.e. when Condition \ref{Condition_trueModel} hold.
To this end, we suppose that the $k^{th}$ element in $X\left(t\right)$
can be written as $X_{k}\left(t\right)=f_{0,k}\left(W\left(t\right),S_{t}N\right)$
where $f_{0,k}:\mathbb{R}^{l}\times\mathbb{M}\rightarrow\mathbb{R}$
such that for any $x\in\mathbb{R}^{l}$, $f_{0,k}\left(x,\cdot\right)$
is nondecreasing and causal, $k=1,2,...,K$. Then, $g_{0}\left(t\right)=f_{0}\left(W\left(t\right),S_{t}N\right)=\sum_{k=1}^{K}b_{k}f_{0,k}\left(W\left(t\right),S_{t}N\right)$
satisfies Condition \ref{Condition_endogeneity} because $b_{k}\geq0$,
$k=1,2,...,K$. The exact functional form of $f_{0,k}$ is not important
for our purposes. However, the following is a simple example of the
monotonicity and independence condition: $f_{0,k}\left(W\left(t\right),S_{t}N\right)=\max\left\{ \min\left\{ 1,W\left(t\right)+\nu\left(S_{t}N\right)\right\} ,0\right\} $
where $\nu\left(S_{t}N\right):=\int_{\left(t-v,t\right)}dN\left(s\right)$
for some finite constant $v>0$, and $S_{t}N$ and $W\left(t\right)$
are independent. Here, $\nu\left(S_{t}N\right)$ counts the most recent
number of events. Recall that we are assuming covariates in $\left[0,1\right]$,
hence $X_{k}\left(t\right):=f_{0,k}\left(W\left(t\right),S_{t}N\right)$
is bounded accordingly.

Condition \ref{Condition_endogeneity} says that we need to be able
to decompose $g_{0}\left(t\right)$ into a part that depends positively
(non-negatively) on past event arrivals (monotonicity) and a component
independent of these. Hence, Condition \ref{Condition_endogeneity}
rules out the case where the impact of $X_{k}\left(t\right)b_{0,k}$
on the current intensity is positive, but the impact of past events
on $X_{k}\left(t\right)b_{0,k}$ is negative. This does not mean that
quantities in the order book cannot reduce the intensity, as this
is just a parametrization problem which is discussed in the remarks
to Condition \ref{Condition_trueModel}. Next we give an example.

For simplicity, with no loss of generality, let $K=1$, so that $X\left(t\right)'b_{0}=X_{1}\left(t\right)b_{0,1}$.
Let $X_{1}\left(t\right)=1-F\left({\rm Dur}\left(t\right)\right)$
where ${\rm Dur}\left(t\right)$ is the last trade duration at time
$t$ and $F\left(\cdot\right)$ is the distribution function of this
duration. Then, for $b_{0,1}>0$ we have that a longer duration implies
a smaller $X_{1}\left(t\right)b_{0,1}$. This makes sense, as a longer
duration implies a smaller intensity. Moreover, it is reasonable that
the next duration is likely to be smaller when past event arrivals
are large. It follows that $X_{1}\left(t\right)b_{0,1}$ is increasing
in $\left(N\left(s\right)\right)_{s<t}$ so that Condition \ref{Condition_endogeneity}
is satisfied. This argument can be extended for $K>1$ to other variables
like spread and order book volume imbalances in exactly the same way.

In the extreme case where $X_{1}\left(t\right)$ is independent of
past event arrivals $\left(N\left(s\right)\right)_{s<t}$, we have
that $X_{1}\left(t\right)$ is strictly exogenous. Then, $X_{1}\left(t\right)b_{0,1}$
can have an impact on $N\left(t\right)$ and future event arrivals
through the intensity $\lambda\left(t\right)$, but is not affected
by past event arrivals $\left(N\left(s\right)\right)_{s<t}$. Given
that we do not assume independence of $\left(N\left(s\right)\right)_{s<t}$,
it is clear why we regard Condition \ref{Condition_endogeneity} to
be a weak exogeneity condition.

\paragraph{Condition \ref{Condition_trueModel}.}

The model is parametrized in a way that is simple and intuitive to
analyze. In practice we will have some raw covariates, say $Z:=\left(Z\left(t\right)\right)_{t\geq0}$
with values in $\mathbb{R}^{K}$ and map them into $\left[0,1\right]^{K}$.
This is always possible because the extended real line is isomorphic
to the unit interval. Such map can change the interpretation of the
covariates. A notable example of such transformation is $F\left(Z_{k}\left(t\right)\right)$
where $F\left(x\right)=\Pr\left(Z_{k}\left(t\right)\leq x\right)$
which can be approximated by the empirical distribution when the variables
are stationary and ergodic. If the covariates take values in a known
compact interval inside the real line, we can just use a linear transformation. 

The raw covariates $Z$ may actually take values in $\mathbb{R}^{L}$
where $L<K$. This is common in many applications where we map the
data into a higher dimensional space in order to capture nonlinearities.
Notable examples are reproducing kernel Hilbert spaces, one-hot encoding
and Bernstein polynomials. In the application of this paper we focus
on one-hot encoding because of their simplicity; a precise definition
of one-hot encoding will be given in due course. Ideally, we would
parametric the model in a parsimonious way. However, ex ante we may
not have a good understanding of how the variables impact the intensity.
Hence, the approach can be seen as either nonparametric or the initial
stage in the analysis of a high frequency dataset. 

Given that we can map the raw variables into a higher dimensional
space, the non-negativity restriction is a parametrization assumption.
In fact, for the one-hot encoding method used in the empirical section,
we have that variables can have a negative impact as soon as the linear
coefficients are decreasing (see Figure \ref{Figure_CLDur98} in Section
\ref{Section_results}). Similar comments pertains to Bernstein polynomials
(e.g. Sancetta, 2018, Section 3.6.6, Mucciante and Sancetta, 2022,
Section 2.2). For the sake of clarity, we now consider two examples. 

Consider the intensity $\lambda\left(Z\left(t\right)\right)=a_{0}+a_{1}Z_{1}\left(t\right)-a_{2}Z_{2}\left(t\right)$
that depends on the raw covariate $Z\left(t\right)=\left[Z_{1}\left(t\right),Z_{2}\left(t\right)\right]'$
with values in $\left[0,1\right]^{2}$, where $a_{i}\geq0$, $i=0,1,2$,
$a_{0}-a_{2}\geq0$. The parameters' restriction ensures that this
intensity is always nonnegative. Then, $\lambda\left(Z\left(t\right)\right)$
can be written as (\ref{EQ_g0Environment}) where $X_{1}\left(t\right)=1$,
$X_{2}\left(t\right)=Z_{1}\left(t\right)$, $X_{3}\left(t\right)=1-Z_{2}\left(t\right)$,
and $b_{0,1}=a_{0}-a_{2}$, $b_{0,2}=a_{1}$, $b_{0,3}=a_{2}$ where
$b_{0}\geq0$. Hence, in our framework we are able to control the
direction of the impact by the linear transformation $x\mapsto1-x$.
From a computational point of view, this is equivalent to changing
the sign of the covariate and imposing an additional inequality constraint.
This example also makes clear that a purely linear model for (\ref{EQ_g0Environment})
is only possible if the raw covariates $Z$ are bounded and we impose
constraints in the estimation. On the other hand, the use of unbounded
variables can lead to a negative intensity.

As a second example we consider the case where each covariate is mapped
into a higher dimensional space by one-hot encoding. For ease of exposition,
consider a model with only one explanatory variable $Z=\left(Z\left(t\right)\right)_{t\geq0}$
, where $Z$ is a predictable ergodic stochastic process such that
$Z\left(t\right)$ takes values in $\mathcal{Z}$, a subset of the
real line, for all $t\geq0$. Let $\mathcal{P}=\bigcup_{k=1}^{K}\mathcal{P}_{k}$
be a partition of $\mathcal{Z}$. One-hot encoding maps the univariate
raw covariate $Z\left(t\right)$ into a $K$ dimensional covariate
$X\left(t\right)$, where the $k^{th}$ element in $X\left(t\right)$
is $X^{\left(k\right)}\left(t\right)=1_{\left\{ Z\left(t\right)\in\mathcal{P}_{k}\right\} }$.
Then, $X$ is said to be a one-hot encoding of $Z$. It follows that
$g_{0}\left(t\right)=\sum_{k=1}^{K}b_{0,k}1_{\left\{ Z\left(t\right)\in\mathcal{P}_{k}\right\} }$
so that the coefficient $b_{0,k}$ is the value of $g_{0}\left(t\right)$
when $Z\left(t\right)\in\mathcal{P}_{k}$. As a concrete illustration,
consider trade durations, where time is measured in seconds. Again,
for simplicity of notation, suppose that the model only uses trade
durations as explanatory variables so that there is only one raw covariate.
In this case $\mathcal{Z}=\left[0,\infty\right)$ because durations
are nonnegative. For a very liquid instrument, we could set $\mathcal{P}_{k}:=\left[\frac{\left(k-1\right)}{10},\frac{k}{10}\right)$,
$k=1,2,...,10$, and $\mathcal{P}_{11}:=\left[1,\infty\right)$. This
means that the univariate raw covariate $Z\left(t\right)$ is mapped
into an 11 dimensional covariate $X\left(t\right)$, so that $K=11$.
In this case if the last duration at time $t$ is 50 milliseconds,
$X^{\left(1\right)}\left(t\right)=1$ and $X^{\left(k\right)}\left(t\right)=0$
for $k\neq1$. This is because 50 milliseconds is 0.05 of a second.
Hence, the duration of 50 milliseconds falls within the first bin
$\mathcal{P}_{1}=\left[0,0.1\right)$. On the other hand if the duration
at time $t$ is 2.5 seconds, we have $X^{\left(11\right)}\left(t\right)=1$
and $X^{\left(k\right)}\left(t\right)=0$ for $k\neq11$. Hence, if
the impact of the duration $Z\left(t\right)$ is monotonically decreasing,
we shall have that $b_{0,k}\geq b_{0,k+1}$ as in fact is the case
for the estimated coefficients in Figure \ref{Figure_CLDur98}. In
our empirical application we use slightly different bins for the durations
and other covariate and the details are reported in Section \ref{Section_oneHotEncodingEmpiricalDetails},
in the Appendix.

The extension to more than one raw covariate is achieved by applying
the one-hot encoding to each variable separately. Suppose that $Z\left(t\right)$
is a $K_{Z}$ dimensional vector where $K_{Z}>1$. The $j^{th}$ element
in $Z\left(t\right)$ is mapped into a $K_{j}$ dimensional covariate
via one-hot encoding. Then, the resulting number of covariates $X$
is $K=\sum_{j=1}^{K_{Z}}K_{j}$.

\paragraph{Condition \ref{Condition_g}.}

The bound $\left|b\right|_{1}\leq B$ is used for technical reasons
to ensure that we achieve consistency under additional regularity
conditions (e.g. Sancetta, 2016, 2018). This is equivalent to estimation
via Lasso because by duality a penalty on the $l_{1}$ norm of the
linear coefficients is equivalent to a constraint on the $l_{1}$
norm of these coefficients. However, given the nonnegativity constraint,
the actual value of $B$ becomes less relevant. The nonnegativity
constraint tends to produce sparse solutions and has some set identification
properties under certain assumptions (Mucciante and Sancetta, 2022).
Using (\ref{EQ_g0Environment}), the $L_{2}$ norm of $\lambda_{0}$
in (\ref{EQ_intensityRepresentation}) can be written as $b'Ab$ where
$A=\frac{1}{T}\int_{0}^{T}\left(h_{0}^{2}\left(t\right)X\left(t\right)X\left(t\right)'\right)dt$
is a $K\times K$ matrix. Then, if for any $b\geq0$ (elementwise)
$b'Ab\geq\nu\left|b\right|_{1}^{2}$ for some $\nu>0$ we have that
a bound on $b'Ab$ controls $\left|b\right|_{1}^{2}$. If this is
the case, intuitively, the sign constraint allows us to control the
$l_{1}$ norm as in Lasso, but through the control of the square error
loss. This condition holds in numerous circumstances (Mucciante and
Sancetta, 2022, Section 2.3). To see this note that all the entries
in $A$ are nonnegative. If we suppose that they are strictly positive,
the inequality $b'Ab\geq\nu\left|b\right|_{1}^{2}$ is trivially satisfied
for all $b\geq0$. Mucciante and Sancetta (2022, Section 2.3) discusses
how the inequality can hold when $A$ is not necessarily strictly
positive. 

\paragraph{Condition \ref{Condition_h}.}

A sum of exponential kernels is a simple but flexible parametrization
and was originally discussed in Hawkes (1971). We choose it for the
sake of definiteness. This allows us to avoid abstract technical conditions.
Any parametric kernel that decays fast enough and is smooth in the
parameters can be used. In this case, consistency can be proved following
the steps in the proofs. Finally, the restriction on the parameter
space to a compact set ensures non-negativity of the intensity and
consistent estimation. 

\subsection{Stationarity and Ergodicity of the Point Process \label{Section_stabilityMultiplicativeHawkes}}

For consistent estimation of the process, we shall use stationarity
and ergodicity. Such statistical properties are of interest in their
own merit and do not need the full extent of the Regularity Conditions.
Hence, we state the following. 

\begin{condition}\label{Condition_minimalRepresentation}There is
a filtration $\mathcal{F}=\left(\mathcal{F}_{t}\right)$ such that
$\lambda_{0}\left(t\right)=h_{0}\left(t\right)g_{0}\left(t\right)$
is an $\mathcal{F}_{t}$-intensity for $N\left(t\right)$ where $h_{0}$
is as in (\ref{EQ_hawkes_ht}) and $g_{0}$ is a uniformly bounded
nonnegative stochastic process satisfying Condition \ref{Condition_endogeneity}
and such that $\left|g_{0}\right|_{\infty}\sum_{l=1}^{L}\frac{d_{0,l}}{a_{0,l}}<1$.\end{condition}

When $g_{0}$ is constant, as in the case of a standard Hawkes process,
Condition \ref{Condition_minimalRepresentation} is the usual one
for stationarity and ergodicity of the Hawkes process (Br\'{e}maud
and Massouli\'{e}, 1996). The uniform upper bound on $g_{0}$ ensures
that we can still apply those results together with the help of Condition
\ref{Condition_endogeneity}. When $g_{0}$ is not constant, Condition
\ref{Condition_minimalRepresentation} restricts $g_{0}$ to be weakly
exogenous in the sense of Condition \ref{Condition_endogeneity}. 

Under the above condition, the point process is strictly stationary
and ergodic. Recall that a point process is stationarity if $N\left(t+C\right)$
and $N\left(C\right)$ have same distribution for any $t\in\mathbb{R}$
and $C\subset\mathbb{R}$. We say that the stationary point process
$N$ is ergodic if $T^{-1}N\left(T\right)\rightarrow\mathbb{E}\lambda_{0}\left(0\right)$
in probability; $\lambda_{0}\left(0\right)$ is $\lambda_{0}\left(t\right)$
for $t=0$. 

We have the following.

\begin{theorem}\label{Theorem_stationarityErgodicity} Suppose that
Condition \ref{Condition_endogeneity} hold.
\begin{enumerate}
\item Then, there is a unique stationary distribution of $N$ with finite
average intensity $\mathbb{E}\int_{0}^{1}dN\left(t\right)$ and dynamics
as in (\ref{EQ_intensityRepresentation}) and the process is ergodic.
\item Suppose that for $t\leq0$, $N$ is restricted to the set $\mathcal{A}:=\left\{ N\left(t\right)=0:t\leq0\right\} $
so that $h_{0}\left(t\right):=c_{0}+\int_{\left(0,t\right)}\sum_{l=1}^{L}d_{0,l}e^{-a_{0,l}\left(t-s\right)}dN\left(s\right)$.
Then, there is a stationary point process $\tilde{N}$ with same dynamics
as in (\ref{EQ_intensityRepresentation}), and a stochastic time $\tau$
such that $\Pr\left(\tau<\infty\right)=1$ and $\tilde{N}\left(t\right)=N\left(t\right)$
for all $t\geq\tau$.
\end{enumerate}
\end{theorem}

The processes $\tilde{N}$ and $N$ couple in finite time irrespective
of the initial condition. This is important because in practice we
only have data for $t>0$. This is equivalent to restricting $N$
to $\mathcal{A}$.

\subsection{Reduced Form Model for Buy and Sell Events\label{Section_reducedFormBuySell}}

Multivariate extensions of intensity models to buy and sell events
have been considered in the literature, but to keep focus we limit
the discussion to univariate intensity (Bauwens and Hautsch, 2009,
and the references in Section \ref{Section_additionalIntroRemarks}
for such extensions). Nevertheless, we show how we can separately
estimate a reduced form model for buy and sell event arrivals. Let
$g^{buy}$ and $g^{sell}$ be as in (\ref{EQ_g0Environment}) but
for buy and sell events separately. Both can depend on order book
variables as well as past durations and other quantities. Suppose
the following feedback loop effect through $g^{buy}$ and $g^{sell}$,
\begin{align}
\lambda^{buy}\left(t\right)= & h^{buy}\left(t\right)\left(g^{buy}\left(t\right)+\rho^{buy}g^{sell}\left(t\right)\right)\nonumber \\
\lambda^{sell}\left(t\right)= & h^{sell}\left(t\right)\left(g^{sell}\left(t\right)+\rho^{sell}g^{buy}\left(t\right)\right),\label{EQ_structuralModel}
\end{align}
where $\rho^{buy}$ and $\rho^{sell}$ are constants in $\left[0,1\right)$.
Because of (\ref{EQ_g0Environment}), this system has the reduced
form 
\begin{align*}
\lambda^{buy}\left(t\right)=h^{buy}\left(t\right) & X\left(t\right)'\left(b^{buy}+\rho^{buy}b^{sell}\right)\left(1-\rho^{buy}\rho^{sell}\right)^{-1}\\
\lambda^{sell}\left(t\right)=h^{sell}\left(t\right) & X\left(t\right)'\left(b^{sell}+\rho^{sell}b^{buy}\right)\left(1-\rho^{buy}\rho^{sell}\right)^{-1}.
\end{align*}
In consequence, separate estimation of the buy and sell intensities
is equivalent to estimation of the above reduced form model as opposed
to (\ref{EQ_structuralModel}). For large samples, the loss in efficiency
for carrying out a separate estimation is secondary.

A natural extension of the multivariate Hawkes process considered
in the literature (Hawkes, 1971) would be $\lambda^{buy}\left(t\right)=h^{buy}\left(t\right)g^{buy}\left(t\right)+\rho^{buy}h^{sell}\left(t\right)g^{sell}\left(t\right)$
and similarly for the sell intensity. This differs from (\ref{EQ_structuralModel}).
In (\ref{EQ_structuralModel}) we are assuming that the intensity
for buy events is independent of $h^{sell}\left(t\right)$ when we
condition on $g^{sell}\left(t\right)$ and $g^{buy}\left(t\right)$,
and similarly for sell events. 

\section{Estimation\label{Section_estimation}}

Throughout, to keep the notation more compact we may write $\int_{0}^{T}fd\mu=\int_{0}^{T}f\left(t\right)dt$
for any measurable function, where $\mu$ is the Lebesgue measure.
The loglikelihood for $\lambda=hg$ as in (\ref{EQ_intensityRepresentation}),
is 
\begin{equation}
L_{T}\left(h,g\right):=\int_{0}^{T}\ln\left(hg\right)dN-\int_{0}^{T}hgd\mu.\label{EQ_logLik}
\end{equation}
Estimation of $g=X'b$ requires a positivity constraint on $b$. This,
coupled with the high dimension $K$ of $b$, makes the problem unfeasible.
Moreover, when the sample size is large, it is not possible to hold
the data in memory. To see this, suppose that $X\left(t\right)=X\left(t_{j}\right)$
for $t\in(t_{j},t_{j+1}]$ where the times $t_{j}$ are update times
for any of the covariates including the counting process, $j=1,2,...,m$.
Throughout, $m$ is the total number of event updates, including the
jumps of the process $N$. The second term in the loglikelihood (\ref{EQ_logLik})
is explicitly written as $\sum_{j=1}^{m}\left(\int_{t_{j}}^{t_{j+1}}h\left(t\right)dt\right)X\left(t_{j}\right)'b$.
This requires to hold in memory a matrix $m\times K$. It is only
feasible for small scale high frequency applications. Even for moderate
scale problems, when $K$ is in the order of hundreds or thousands,
and $m$ is a few million (e.g. a few days for some liquid instruments),
estimation of $g$ by loglikelihood with a positivity constraint is
impractical. We may have only a few thousand events generated by $N$
in a day, but the order book updates much faster and we wish to track
the impact of order book derived covariates. 

It is simple to see that around the true value the negative loglikelihood
is quadratic. In fact, an alternative estimation strategy for intensity
processes is based on least squares (Ga\"{i}ffas and Guilloux, 2012,
Mucciante and Sancetta, 2022, and references therein). As a first
step, we suggest to minimize the quadratic loss 
\begin{equation}
Q_{T}\left(h,g\right):=-\frac{2}{T}\int_{0}^{T}hgdN+\frac{1}{T}\int_{0}^{T}\left(hg\right)^{2}d\mu.\label{EQ_quadraticRisk}
\end{equation}
When, $h$ is known, minimization is just a quadratic programming
problem, hence much easier to solve than (\ref{EQ_logLik}). The rationale
for minimizing (\ref{EQ_quadraticRisk}) is that its expectation is
equal to 
\[
\mathbb{E}Q_{T}\left(h,g\right):=-\frac{2}{T}\mathbb{E}\int_{0}^{T}hg\lambda_{0}d\mu+\frac{1}{T}\mathbb{E}\int_{0}^{T}\left(hg\right)^{2}d\mu
\]
and it is minimized by $h=h_{0}$ and $g=g_{0}$ because $\lambda_{0}=h_{0}g_{0}$.
When $h$ is unknown, the second term in (\ref{EQ_quadraticRisk})
poses similar computational challenges as the second term in (\ref{EQ_logLik})
for large datasets. To solve this, we add a second step in the optimization
procedure. We suggest to fix $g$ and estimate $h$ and vice versa,
as commonly done in coordinate descent algorithms. When it comes to
optimization w.r.t. $g$, given an $h$, we use the loss function
\begin{equation}
R_{T}\left(g;h\right):=-\frac{2}{T}\int_{0}^{T}\frac{g}{h}dN+\frac{1}{T}\int_{0}^{T}g^{2}d\mu\label{EQ_contrastFunctionUnfeasible-1}
\end{equation}
in place of (\ref{EQ_quadraticRisk}). This is a much simpler problem,
as there is a summary statistic for the second term as in standard
regression problems. In summary, we alternate between estimation of
$h$ minimizing (\ref{EQ_quadraticRisk}) with $g$ fixed and estimation
of $g$ minimizing (\ref{EQ_contrastFunctionUnfeasible-1}) with $h$
fixed. The starting value for $g$ is set to a constant so that at
the first iteration we estimate a standard Hawkes process using (\ref{EQ_quadraticRisk}).
Algorithm \ref{Algorithm_1} summarizes the procedure. There, the
only free parameter is $B$, as defined in Condition 2. We discuss
its choice in Section \ref{Section_choiceB}.

The theory of coordinate descent algorithms justifies alternating
between minimization w.r.t. $g$ and $h$ (Beck and Tetruashvili,
2013). We shall show that asymptotically, the minimizers of (\ref{EQ_quadraticRisk})
and (\ref{EQ_contrastFunctionUnfeasible-1}) w.r.t. $g\in\mathcal{G}$
are the same if $h$ is close to $h_{0}$.

Estimation via the loss function (\ref{EQ_contrastFunctionUnfeasible-1})
is very efficient as it is a standard quadratic programming problem.
Let $T_{j}$ be the $j^{th}$ jump time of $N$, $j=1,2,...,n$ with
$n=N\left(T\right)$ and for simplicity suppose that $t_{m}=T$. For
numerical estimation we rewrite (\ref{EQ_contrastFunctionUnfeasible-1})
explicitly as 
\begin{equation}
R_{T}\left(b\right):=-\frac{2}{T}b'\Phi'\Gamma+\frac{1}{T}b'\Phi'\Sigma\Phi b\label{EQ_contrastMatrixNotation}
\end{equation}
where the $j^{th}$ row of $\Phi$ is $X\left(t_{j}\right)$, $\Gamma$
is a vector with $j^{th}$ entry $1/h(T_{l})$ if $t_{j}=T_{l}$ for
some $l$ (i.e., if $t_{j}$ is a jump time of $N$) and zero otherwise;
$\Sigma$ is a diagonal matrix with $\left(j,j\right)^{th}$ entry
$\left(t_{j}-t_{j-1}\right)$. Note that when a new observation is
collected, we only need to update $\Phi'\Gamma$ and $\Phi'\Sigma\Phi$,
which are low dimensional matrices ($K\times1$ and $K\times K$,
respectively) relatively to $m$. Moreover, the dimension of $\Phi'\Sigma\Phi$
does not depend on $m$ (very large) and on $h$. 

We may incur additional difficulties when the $K$-dimensional process
$X$ has linearly dependent entries. This is the case for our application
based on one-hot encoding as well as for other methods such as estimation
in reproducing kernel Hilbert spaces or additive multivariate Bernstein
polynomials. However, the nonnegativity constraints mitigates this
problem if we can find a $\nu>0$ such that $b'\left(\Phi'\Sigma\Phi/T\right)b\geq\nu b'b$
for all $b\geq0$, where the inequality on $b$ is meant elementwise
(see the discussion to Condition \ref{Condition_g} in Section \ref{Section_remarksRegularityConditions}).
In the applications we have considered the problem remained convex
and a fast solution could be obtained. For more difficult problems,
we may rely on greedy algorithms (e.g. Sancetta, 2016, 2018, and references
therein).

When the number of updates is very large, computation of the second
integral in (\ref{EQ_quadraticRisk}) can be slow even for fixed $g$.
Monte Carlo techniques can be used in this case (Cartea et al., 2021). 

\begin{algorithm}[H]
\caption{Intensity Estimation}
\label{Algorithm_1}

Start with $g^{\left(0\right)}=\gamma$. 

For each $v=1,2,...$, find the minimizer of $Q_{T}\left(h,g^{\left(v-1\right)}\right)$
(as in (\ref{EQ_quadraticRisk})) w.r.t. $h$ and denote it by $h^{\left(v\right)}$.
When $v=1$ we shall also minimize w.r.t. $\gamma>0$.

Minimize $R_{T}\left(X'b;h\right)$ (as in (\ref{EQ_contrastFunctionUnfeasible-1}))
w.r.t. $b\in\left[0,\infty\right)^{K}$ s.t. $\sum_{k}b_{k}\leq B$. 

Define the minimizer by $b^{\left(v\right)}$ so that $g^{\left(v\right)}=X'b^{\left(v\right)}$
is the estimator for $g_{0}$.

Stop when $h^{\left(v\right)}g^{\left(v\right)}$ converges.
\end{algorithm}

\paragraph{Reparametrization for improved estimation.}

We found improvements in the estimation of (\ref{EQ_intensityRepresentation})
using the reparametrization 
\begin{equation}
\lambda_{0}\left(t\right)=\left(c_{0}d_{0}+\int_{\left(-\infty,t\right)}\sum_{l=1}^{L}d_{0,l}e^{-a_{0,l}\left(t-s\right)}dN\left(s\right)\right)\frac{g_{0}\left(t\right)}{\mathbb{E}g_{0}}.\label{EQ_modelReparametrization}
\end{equation}
In (\ref{EQ_quadraticRisk}), we would estimate $c$ and $d$ using
$g/\mathbb{E}g$ as given, instead of $g$. We would then estimate
$g$ in (\ref{EQ_contrastFunctionUnfeasible-1}) using $h\left(t\right)=\left(c+\int_{\left(-\infty,t\right)}\sum_{l=1}^{L}\frac{d_{l}}{d_{1}}e^{-a_{0,l}\left(t-s\right)}dN\left(s\right)\right)$
as given; for identification, this forces $d_{l}/d_{1}=1$ when $l=1$.
This means that at every iteration $\mathbb{E}g$ is replaced with
the most recent estimator for $\mathbb{E}g$, for example, in the
empirical section, at the $v^{th}$ iteration we use $\mathbb{E}g\simeq\frac{1}{m}\sum_{j=1}^{m}X\left(t_{j}\right)'b^{\left(v-1\right)}$;
the update times $t_{j}$ were defined below (\ref{EQ_logLik}). 

\subsection{Choice of $B$\label{Section_choiceB}}

The non-negativity constraint leads to the shrinkage property studied
in Mucciante and Sancetta (2022), among others. Therefore, the choice
of $B$ (Condition \ref{Condition_g}) is not very important. 

Nevertheless, following Sancetta (2018) the value $B$ can be chosen
to minimize $-L_{T}\left(\hat{h}_{B},\hat{g}_{B}\right)+K_{B}$ w.r.t.
$B$, where $L_{T}$ is the loglikelihood in (\ref{EQ_logLik}) and
$\hat{h}_{B},\hat{g}_{B}$ are the estimators from Algorithm \ref{Algorithm_1}
for a given $B$. Here, $K_{B}$ is the number of nonzero coefficient
in the estimator for $b_{0}$. We note that even when $B=\infty$,
the number $K_{B}$ is smaller than $K$, which is the size of $b_{0}$.
This is due to the aforementioned shrinkage property.

To identify a reasonable order of magnitude for $B$, we can use the
fact that $\mathbb{E}h_{0}g_{0}\leq\mathbb{E}h_{0}\left|g_{0}\right|_{\infty}=\left[c_{0}/\left(1-\sum_{l=1}^{L}\frac{d_{0,l}}{a_{0,l}}\right)\right]B$.
Moreover, $\mathbb{E}h_{0}g_{0}$ is approximately $N\left(T\right)/T$
so that we would expect $B$ to be greater than $\left[c_{0}/\left(1-\sum_{l=1}^{L}\frac{d_{0,l}}{a_{0,l}}\right)\right]^{-1}N\left(T\right)/T$.
In practice, we replace $c_{0}$, $d_{0,l}$ and $a_{0,l}$ with the
estimated parameters from a Hawkes process with $g_{0}$ constant.
We can also approximate $B$ to be of the same order of magnitude
as $d_{0,1}$ when we use the parametrization in (\ref{EQ_modelReparametrization}).
Assuming that on average the covariates have expectation $1/2$, then
we would have $B=2d_{0,1}$. The simulation results show that such
large value is reasonable and results are not sensitive to it.

A related approach is to directly impose a constraint on the magnitude
of each coefficient. Assume for the moment that $b=\beta1_{K}$ for
some constant $\beta>0$, and solve (\ref{EQ_contrastMatrixNotation})
under this constraint. The solution is $\beta=1_{K}'\Phi'\Gamma/\left(1_{K}'\Phi'\Sigma\Phi1_{K}\right)$.
We can then solve a quadratic programming problem under the constraint
that $b_{k}\in\left[0,\beta\right]$, which clearly implies $B=K\beta$.
We chose such approach in the empirical section and found our results
not to be sensitive to the above alternatives. 

\subsection{Simulations: Number of Iterations for the Algorithm to Converge\label{Section_simulationsConvergence} }

We found that the algorithm would converge within two or three iterations.
We showcase this with a simulation. We let $X\left(t\right)=X\left(T_{i-1}\right)$
when $t\in\left(T_{i-1},T_{i}\right]$ and let $X\left(T_{i-1}\right)$
be uniformly distributed in $\left[0,1\right]^{K}$ for $i\geq0$.
Moreover, we set $b_{0}$ such that the first three entries are equal
to $2/3$ and the remaining are equal to zero. This choice implies
that $\mathbb{E}g=1$. Moreover, we consider $L=1$ and $\left(c_{0},d_{0},a_{0}\right)=\left(1,1,2\right)$,
dropping the subscript $l$. We use Algorithm \ref{Algorithm_1} for
the estimation. We use the following parameters restrictions: $B=K$
and $a\in\left[10^{-9},10^{4}\right]$, $c\in\left[10^{-9},10\right]$,
$d\in\left[10^{-9},10^{3}\right]$. We cannot report all estimated
values of $b$. To help assess how close estimated parameters are
to the true ones, we compute
\begin{equation}
{\rm Error}\left(\alpha\right):=\sum_{k=1}^{K}1_{\left\{ \left|b_{k}-b_{0,k}\right|>\alpha\left(2/3\right)\right\} }\label{EQ_errorBHatB0}
\end{equation}
where $2/3$ is the value of the non-zero entries in $b_{0}$ and
$\alpha$ is a small number. We compute the above quantities from
a single realization of 100,000 jumps from the process. As we increase
the number of iterations, the convergence is remarkably fast. To reduce
the dependence on a the single sample realization, we also compute
an average, at the fourth iteration, of the end results and their
standard errors over 50 realization. The average numbers over the
simulations confirm that the claim is not just the result of chance.
Table \ref{Table_algoConvergence} reports the details. From those
results we deduce that the error in the estimation is relatively low
(see columns $\alpha=0.1,0.05$ in Table \ref{Table_algoConvergence}).
We can deduce that two/three iterations are enough to converge. Moreover,
the results in the row marked as ``avg'' of Table \ref{Table_algoConvergence}
show that the procedure has good finite sample properties. In particular,
for this simulation design, we verify the variable screening properties
of the methodology. Mutatis mutandis, simulations in Mucciante and
Sancetta (2022) shed additional light in this respect, in a richer
variety of simulation designs when $h_{0}=1$. 

\begin{table}
\caption{Example of Algorithm Performance. Estimation is for $b\in\left[0,B\right]^{K}$
with $B=K$. The true values are $b_{0,k}=2/3$ for $k=1,2,3$ and
zero otherwise, and $\left(c_{0},d_{0},a_{0}\right)=\left(1,1,2\right)$.
Their estimated values are reported in columns $b_{1},b_{2},b_{3}$$,c,d,a$
for one single simulation over increasing number of iterations (iter)
in Algorithm \ref{Algorithm_1}. The average (avg.) and standard error
(s.e.) over 50 simulations are given in the last two rows for iter
4. Columns ${\rm Error}\left(\alpha\right)$ compute the statistic
(\ref{EQ_errorBHatB0}) for different values of $\alpha$. Small values
imply higher precision of the estimator. }
\label{Table_algoConvergence}

\begin{tabular}{cccccccccccc}
 &  &  &  &  &  &  &  &  &  &  & \tabularnewline
 &  &  &  &  &  &  &  & \multicolumn{3}{c}{${\rm Error}\left(\alpha\right)$ in (\ref{EQ_errorBHatB0})} & \tabularnewline
 &  &  &  &  &  &  &  & \multicolumn{3}{c}{$\alpha$} & \tabularnewline
$K$ & iter & $b_{1}$ & $b_{2}$ & $b_{3}$ & $c$ & $d$ & $a$ & $0.1$ & $0.05$ & $0.01$ & \tabularnewline
\cline{1-11}
3 & 1 & 0.675 & 0.640 & 0.684 & 0.701 & 1.130 & 2.171 & 0 & 0 & 3 & \tabularnewline
 & 2 & 0.677 & 0.637 & 0.684 & 1.014 & 0.988 & 2.008 & 0 & 0 & 3 & \tabularnewline
 & 3 & 0.677 & 0.637 & 0.684 & 1.014 & 0.988 & 2.008 & 0 & 0 & 3 & \tabularnewline
 & 4 & 0.677 & 0.637 & 0.684 & 1.014 & 0.988 & 2.008 & 0 & 0 & 3 & \tabularnewline
 & avg & 0.668 & 0.667 & 0.665 & 0.998 & 0.999 & 2.000 & 0 & 0 & 1.260 & \tabularnewline
 & s.e. & 0.000 & 0.000 & 0.000 & 0.000 & 0.000 & 0.000 & 0.000 & 0.000 & 0.112 & \tabularnewline
10 & 1 & 0.648 & 0.614 & 0.658 & 0.701 & 1.130 & 2.171 & 0 & 1 & 8 & \tabularnewline
 & 2 & 0.677 & 0.637 & 0.684 & 1.007 & 0.997 & 2.013 & 0 & 0 & 3 & \tabularnewline
 & 3 & 0.677 & 0.637 & 0.684 & 1.014 & 0.988 & 2.008 & 0 & 0 & 3 & \tabularnewline
 & 4 & 0.677 & 0.637 & 0.684 & 1.014 & 0.988 & 2.008 & 0 & 0 & 3 & \tabularnewline
 & avg & 0.664 & 0.663 & 0.661 & 0.998 & 1.000 & 2.001 & 0.000 & 0.000 & 2.300 & \tabularnewline
 & s.e. & 0.000 & 0.000 & 0.000 & 0.000 & 0.000 & 0.000 & 0.000 & 0.000 & 0.188 & \tabularnewline
100 & 1 & 0.635 & 0.600 & 0.645 & 0.701 & 1.130 & 2.171 & 0 & 1 & 10 & \tabularnewline
 & 2 & 0.674 & 0.634 & 0.681 & 1.005 & 1.002 & 2.019 & 0 & 0 & 3 & \tabularnewline
 & 3 & 0.675 & 0.635 & 0.682 & 1.014 & 0.989 & 2.009 & 0 & 0 & 3 & \tabularnewline
 & 4 & 0.675 & 0.635 & 0.682 & 1.014 & 0.989 & 2.009 & 0 & 0 & 3 & \tabularnewline
 & avg & 0.659 & 0.658 & 0.655 & 0.997 & 1.002 & 2.003 & 0.000 & 0.020 & 3.420 & \tabularnewline
 & s.e. & 0.000 & 0.000 & 0.000 & 0.000 & 0.000 & 0.000 & 0.000 & 0.003 & 0.351 & \tabularnewline
\end{tabular}
\end{table}

The results show that the parameters converge fast with the number
of iterations. The convergence is towards the true parameters. We
also report sensitivity results around different choices of $B$.
In particular, we consider 
\begin{equation}
B={\rm Mult}.\times\left[\hat{c}/\left(1-\hat{a}^{-1}\right)\right]^{-1}N\left(T\right)/T\label{EQ_BChoice}
\end{equation}
 where $\hat{a}$ and $\hat{c}$ are the estimators for $a_{0}$ and
$c_{0}$ at the first iteration in Algorithm \ref{Algorithm_1} and
${\rm Mult.}\in\left\{ 0.5,1,10\right\} $. Note that when ${\rm Mult.}=0.5$,
the resulting $B$ is smaller than the recommendation from Section
\ref{Section_choiceB} and we expect a sub-optimal performance. We
also consider $B=K$ which satisfies the requirement that $B>\mathbb{E}g_{0}$
as discussed in Section \ref{Section_choiceB}. Under various criteria,
we find out that the choice of $B$ is not critical as long as the
recommendations from Section \ref{Section_choiceB} are followed.
The results in Table \ref{Table_BSensitivity} corroborate this statement.

\begin{table}
\caption{Sensitivity of Results to $B$. Results for different metrics are
reported as we vary the value of $B$ using (\ref{EQ_BChoice}). When
a value for ${\rm Mult.}$ is not reported, it means that $B=K$.
The average (avg.) and standard error (s.e.) over 50 simulations are
reported. The $l_{1}$ and $l_{2}$ norms of the difference between
the estimator and the true one are scaled by the norm of the true
parameter. The columns FP, FN and P report the false positive, false
negative and the positives: the number of non-zero estimated coefficients
that should in fact be zero, the number of zero estimated coefficient
that instead should be non zero, and the number of non-zero coefficients
in the population. The estimated value of $B$ is also reported. }
\label{Table_BSensitivity}

\begin{tabular}{cccccccccc}
$K$ & ${\rm Mult.}$ &  & $l_{1}$ & $l_{2}$ & FP & FN & P & estimated $B$ & \tabularnewline
\cline{1-9}
3 & 0.5 & avg & 0.0281 & 0.0334 & 0 & 0 & 3 & 0.6496 & \tabularnewline
 &  & s.e. & 0.0004 & 0.0050 & 0 & 0 & 0 & 0.0003 & \tabularnewline
 & 1 & avg & 0.0150 & 0.0191 & 0 & 0 & 3 & 1.2991 & \tabularnewline
 &  & s.e. & 0.0001 & 0.0026 & 0 & 0 & 0 & 0.0006 & \tabularnewline
 & 10 & avg & 0.0150 & 0.0191 & 0 & 0 & 3 & 12.9915 & \tabularnewline
 &  & s.e. & 0.0001 & 0.0026 & 0 & 0 & 0 & 0.0057 & \tabularnewline
 & - & avg & 0.0150 & 0.0191 & 0 & 0 & 3 & 3 & \tabularnewline
 &  & s.e. & 0.0001 & 0.0026 & 0 & 0 & 0 & 0 & \tabularnewline
10 & 0.5 & avg & 0.0540 & 0.0412 & 7 & 0 & 3 & 0.6496 & \tabularnewline
 &  & s.e. & 0.0005 & 0.0054 & 0 & 0 & 0 & 0.0003 & \tabularnewline
 & 1 & avg & 0.0215 & 0.0221 & 7 & 0 & 3 & 1.2991 & \tabularnewline
 &  & s.e. & 0.0002 & 0.0029 & 0 & 0 & 0 & 0.0006 & \tabularnewline
 & 10 & avg & 0.0215 & 0.0221 & 7 & 0 & 3 & 12.9915 & \tabularnewline
 &  & s.e. & 0.0002 & 0.0029 & 0 & 0 & 0 & 0.0057 & \tabularnewline
 & - & avg & 0.0215 & 0.0221 & 7 & 0 & 3 & 10 & \tabularnewline
 &  & s.e. & 0.0002 & 0.0029 & 0 & 0 & 0 & 0 & \tabularnewline
100 & 0.5 & avg & 0.0585 & 0.0385 & 97 & 0 & 3 & 0.6496 & \tabularnewline
 &  & s.e. & 0.0005 & 0.0051 & 0 & 0 & 0 & 0.0003 & \tabularnewline
 & 1 & avg & 0.0324 & 0.0265 & 97 & 0 & 3 & 1.2991 & \tabularnewline
 &  & s.e. & 0.0003 & 0.0033 & 0 & 0 & 0 & 0.0006 & \tabularnewline
 & 10 & avg & 0.0324 & 0.0265 & 97 & 0 & 3 & 12.9915 & \tabularnewline
 &  & s.e. & 0.0003 & 0.0033 & 0 & 0 & 0 & 0.0057 & \tabularnewline
 &  & avg & 0.0324 & 0.0265 & 97 & 0 & 3 & 100 & \tabularnewline
 &  & s.e. & 0.0003 & 0.0033 & 0 & 0 & 0 & 0 & \tabularnewline
 &  &  &  &  &  &  &  &  & \tabularnewline
\end{tabular}
\end{table}

\section{Asymptotic Results\label{Section_asymptoticResults}}

First we show that estimation of the model using the quadratic loss
(\ref{EQ_quadraticRisk}) gives consistent estimators even in the
ultra high dimensional case. 

\begin{theorem}\label{Theorem_QuadraticLossConsistency} Define $\left(\tilde{h},\tilde{g}\right):=\arg\inf_{h,g}Q_{T}\left(h,g\right)$.
Under the Regularity Conditions, 
\[
\frac{1}{T}\int_{0}^{T}\left(h_{0}g_{0}-\tilde{h}\tilde{g}\right)^{2}d\mu=O_{P}\left(B\sqrt{\frac{\ln\left(1+K\right)}{T}}\right).
\]
\end{theorem}

Define $\hat{g}^{h}=\arg\inf_{g\in\mathcal{G}}R_{T}\left(g;h\right)$
and $g^{h}=\arg\inf_{g\in\mathcal{G}}\frac{1}{T}\int_{0}^{T}\left(\frac{\lambda_{0}}{h}-g\right)^{2}d\mu$.
We show that the minimizer $\hat{g}^{h}$ converges to the best $L_{2}$
approximation $g^{h}$ uniformly in $h$ under certain conditions.
We can then conclude that in a neighbourhood of $h_{0}$ the estimators
of $g_{0}$ using the loss functions (\ref{EQ_quadraticRisk}) and
(\ref{EQ_contrastFunctionUnfeasible-1}) are equivalent. We make this
clear with the following.

\begin{theorem}\label{Theorem_convergence} Under the Regularity
Conditions, 
\begin{enumerate}
\item $\sup_{g\in\mathcal{G}}\left|R_{T}\left(g;h\right)-\frac{1}{T}\int_{0}^{T}\left(\frac{\lambda_{0}}{h}-g\right)^{2}d\mu\right|=\frac{1}{T}\int_{0}^{T}\left(\lambda_{0}/h\right)^{2}d\mu+O_{P}\left(B\sqrt{\frac{\ln\left(1+K\right)}{T}}\right)$
uniformly in $h\in\mathcal{H}$; moreover we have that $\sup_{h\in\mathcal{H}}\frac{1}{T}\int_{0}^{T}\left(\lambda_{0}/h\right)^{2}d\mu\leq\frac{1}{T}\int_{0}^{T}\underline{c}^{-2}\lambda_{0}^{2}d\mu$
where $\underline{c}=\min\left\{ c\in\mathcal{C}\right\} $;
\item for any set $\mathcal{H}'\subseteq\mathcal{H}$, we have that $\sup_{h\in\mathcal{H}'}\frac{1}{T}\int_{0}^{T}\left(\hat{g}^{h}-g^{h}\right)^{2}=O_{P}\left(\max\left\{ B\sqrt{\frac{\ln\left(1+K\right)}{T}},A_{T}\right\} \right)$,
where $A_{T}=\sup_{h\in\mathcal{H}'}\frac{1}{T}\int_{0}^{T}\left(\frac{\lambda_{0}}{h}-g^{h}\right)^{2}d\mu$;
\item $\frac{1}{T}\int_{0}^{T}\left(\hat{g}^{\hat{h}}-g_{0}\right)^{2}=O_{P}\left(B\sqrt{\frac{\ln\left(1+K\right)}{T}}\right)$,
if $\hat{h}\in\mathcal{H}$ is such that $\frac{1}{T}\int_{0}^{T}\left(\hat{h}g_{0}-h_{0}g_{0}\right)^{2}=O_{P}\left(B\sqrt{\frac{\ln\left(1+K\right)}{T}}\right)$.
\end{enumerate}
\end{theorem}

In Point 1, the second term on the right hand side (r.h.s.) does not
depend on $g$ and $h$, hence the objective function converges uniformly
to the average square loss plus some quantity independent of $g$.
Note also that $\int_{0}^{T}\left(\frac{\lambda_{0}}{h}-g\right)^{2}d\mu=\int_{0}^{T}\left(\frac{\lambda_{0}-hg}{h}\right)^{2}d\mu$.
If the approximation of the scaled intensity $\lambda_{0}/h$ by $g^{h}$
is good in the sense that the approximation error $A_{T}$ in Point
2 is small, then the estimator $\hat{g}^{h}$converges to the best
approximation uniformly in $h\in\mathcal{H}'$. For example, $\mathcal{H}'$
could a ball of small radius around $h_{0}$. This is made clearer
in Point 3. If we can find $\hat{h}\rightarrow h_{0}$ in $L_{2}$
in probability, then also $\hat{g}^{h}\rightarrow g_{0}$ in $L_{2}$
in probability. The contribution to the error that comes from the
dimensionality $K$ of the covariates is only logarithmic. Hence,
the methodology is suited for potentially ultra high dimensional models. 

To put the rates of convergence into perspective, note that in the
regression setting with i.i.d. Gaussian errors, no estimator of an
arbitrary convex combinations of variables can achieve a rate faster
than $\left(\frac{\ln K}{n}\right)^{1/4}$, where here $n$ is the
sample size (Tsybakov, 2003, Theorem 2). In our case $n$ can be viewed
as the number of event event arrivals $N\left(T\right)$. Hence, the
result from Theorems \ref{Theorem_QuadraticLossConsistency} and \ref{Theorem_convergence}
can be considered optimal unless we add additional conditions. 

\subsection{Test Statistic to Compare Two Intensity Estimators \label{Section_ComparingIntensities}}

A common diagnostic test for point processes is to use the fact that
$\int_{T_{i-1}}^{T_{i}}\lambda_{0}\left(t\right)dt$, $i=1,2,3,...$
are i.i.d. exponential random variables (Bauwens and Hautsch, 2009,
Section 4.3); recall that the times $T_{i}$ are the times of the
event arrival. Such method cannot be used to compare the relative
fit of two competing intensities. For this reason, methods based on
the loglikelihood ratio are better suited. Due to the fact that the
data is possibly high dimensional, the loglikelihood ratio can have
nonstandard distribution in high dimensions. To avoid this problems
we consider a sample splitting procedure. Then, the result also applies
to the intensity that is recursively estimated. All that we need is
a measurable intensity that is bounded away from zero. Using a sample
up to time $0$, the estimator for two competing intensities $\lambda^{(k)}$,
$k=1,2$, is denoted by $\hat{\lambda}^{(k)}$. This estimator is
evaluated on data on the following days. For model $k$, the log likelihood
(see (\ref{EQ_logLik})) evaluated on a sample $(0,T]$ using $\hat{\lambda}^{(k)}$
is denoted by $L_{T}^{(k)}$. Let $q_{\alpha}$ be the $\alpha$ quantile
of the standard normal distribution, e.g., $q_{95}\simeq1.64$. At
the $\left(1-\frac{\alpha}{100}\right)\%$ significance level for
a one sided test, we reject model $2$ in favor of model 1 if 
\begin{equation}
\frac{L_{T}^{(1)}-L_{T}^{(2)}}{\sqrt{T\hat{\sigma}_{T}^{2}}}\geq q_{\alpha}\label{EQ_testRejectRule}
\end{equation}
where
\begin{equation}
\hat{\sigma}_{T}^{2}=\frac{1}{T}\int_{0}^{T}\big[\ln\left(\hat{\lambda}^{\left(1\right)}/\hat{\lambda}^{(2)}\right)\big]^{2}dN.\label{EQ_sampleVarLogLik}
\end{equation}
The predictable part of the log-likelihood $L_{T}^{\left(k\right)}$
is 
\[
H_{T}^{\left(k\right)}:=\int_{0}^{T}\ln\left(\hat{\lambda}^{\left(k\right)}\right)\lambda_{0}d\mu-\int_{0}^{T}\hat{\lambda}^{\left(k\right)}d\mu.
\]
The predictable part of the likelihood ratio is such that $H_{T}^{\left(k\right)}$
is maximized when $\hat{\lambda}^{\left(j\right)}=\lambda_{0}$. The
closer $\hat{\lambda}^{\left(1\right)}$ is to $\lambda_{0}$ relatively
to $\hat{\lambda}^{\left(2\right)}$ the larger is $H_{T}^{\left(1\right)}-H_{T}^{\left(2\right)}$.
Closeness of $\hat{\lambda}^{\left(k\right)}$ to $\lambda_{0}$ in
measured in the sense of Kulback-Liebler divergence, i.e. how $H_{T}^{\left(0\right)}-H_{T}^{\left(k\right)}\geq0$
is close to zero, where $H_{T}^{\left(0\right)}$ is as in the above
display replacing $\hat{\lambda}^{\left(k\right)}$ with $\lambda_{0}$.
Define the predictable part of the log-likelihood ratio as 
\begin{equation}
\epsilon_{T}:=H_{T}^{\left(1\right)}-H_{T}^{\left(2\right)}.\label{EQ_predictableLogLikRatio}
\end{equation}
Under the null hypothesis, we suppose that $\epsilon_{T}=o_{p}\left(\sqrt{T}\right)$.
Loosely speaking, the intensities $\hat{\lambda}^{\left(1\right)}$
and $\hat{\lambda}^{\left(2\right)}$ give similar predictions asymptotically,
i.e. the predictable part of the log-likelihood ratio diverges at
a rate slower than $o_{p}\left(\sqrt{T}\right)$. This means that
we can only distinguish intensities that are sufficiently far apart
as $T\rightarrow\infty$. The rate of divergence of $\epsilon_{T}$
is clearly of order $T$ when the two intensities differ by some positive
function that is bounded away from zero. The following can be used
to justify (\ref{EQ_testRejectRule}). The next result generalizes
Proposition 1 in Sancetta (2018) to unbounded intensities. 

\begin{theorem}\label{Theorem_testCompetingIntensities}Suppose that
the Regularity Conditions hold, that $\hat{\lambda}^{\left(k\right)}\in\mathcal{H}\times\mathcal{G}$,
$k=1,2$, $\hat{\lambda}^{\left(1\right)}\neq\hat{\lambda}^{\left(2\right)}$
and such that no intensity is exactly zero. If $\epsilon_{T}=o_{p}\left(\sqrt{T}\right)$,
then,
\begin{equation}
\frac{L_{T}^{\left(1\right)}-L_{T}^{\left(2\right)}}{\sqrt{T\hat{\sigma}_{T}^{2}}}\rightarrow Z\label{eq:T1}
\end{equation}
in distribution where $Z$ is a standard normal random variable.\end{theorem}

\section{Empirical Application\label{Section_empirical}}

We separately model trade arrivals on each side of the order book.
It is also of interest to consider trade arrivals for large trade
sizes. This is because for large trade sizes a market maker might
face averse selection. This means being filled on a passive/resting
order when the market goes in the opposite direction of the filled
order. For example, a market maker may have an order to sell at the
top of book ask. A large buy trade is initiated and it depletes the
whole first level on the asks. In consequence, the market maker sells
at the ask, but the price jumps up, making the sale unprofitable,
at least in the short term. In what follows we consider two type of
events: 1. trades arrivals on one side of the book, irrespective of
size, and 2. trades arrivals on one side of the book for sizes that
are at least as large as the size posted on the top of book. In what
follows, we shall refer to these as Any Trade Arrivals and Large Trade
Arrivals. 

In particular, we study four stocks traded on the NYSE: Amazon (AMZN),
Cisco (CSCO), Disney (DIS) and Coca Cola (KO). We also use the ETF
on the S\&P500 (SPY) as auxiliary instrument. The stocks tickers are
given inside the parenthesis. The sample period is 01/March/2019-30/April/2019
from 9:30am until 4:30pm every trading day. 

\paragraph{Objectives.}

We use our flexible methodology to answer a number of questions. 
\begin{enumerate}
\item Does the order book provide information in addition to what is captured
by trades?
\item Is self excitation important once we account for the order book?
\item Is the impact of such information nonlinear?
\item Is the simple exponential kernel ($L=1$ in (\ref{EQ_hawkes_ht}))
enough to capture the self excitation nature of event arrivals?
\end{enumerate}
To do so, we test various functional restrictions within our framework.
We test the hypotheses that $h_{0}$ is constant against the unrestricted
model, and that $g_{0}$ is constant against the unrestricted model.
To account for possible nonlinearities, we use one-hot-encoding to
obtain an estimator of $g_{0}$. To verify whether the nonlinearity
is important, we also allow $g_{0}$ to be linear in the underlying
raw variables, i.e. without mapping them into a higher dimensional
space using one-hot-encoding. In this case, we do not restrict the
coefficients to be nonnegative. The out of sample test procedure for
all these specifications is discussed in Section \ref{Section_ComparingIntensities}. 

\paragraph{Data quality.}

The data was obtained from the Lobster database querying the first
ten levels of the order book. The Lobster dataset is based on Level
3 data. It means that for each given stock, every order is included,
e.g. insertion, cancellation, execution of visible and hidden limit
orders. Lobster construct a snapshot of the order book for any such
orders. Details and nuisances about the dataset can be found in Section
\ref{Section_dataConflationArrivals} of the Appendix. 

\paragraph{Data size.}

We conflate multiple updates into a single one if the exchange time
stamp is the same. Then, for AMZN, we have more that 17 million book
updates with unique time stamp for ten levels and in excess of 400
thousand trades over the whole sample of 42 available trading days
(Table \ref{Table_summaryStats}). We use the first three levels,
and these usually account for about 60\% of the updates. The number
of parameters $K$ to be estimated is at most 177. This means about
300 million data points. Our estimation procedure had no problems
to load the data in RAM, and was rather fast (less than two hours
to parse data, a day at the time, and estimate a model). 

\subsection{The Raw Covariates}

We estimate (\ref{EQ_intensityRepresentation}) where $X$ is a one-hot
encoding of the raw covariates reported in Table \ref{Table_covariatesEmpiricalSection}.
Some of these raw covariates are obtained applying exponential moving
average (EWMA) filters to the data. This is the case if a smoothing
parameter is specified in Table \ref{Table_covariatesEmpiricalSection}.
The EWMA of a variable $Z\left(t_{i}\right)$ with smoothing parameter
$\alpha$ is 
\begin{equation}
EWMA\left(Z\left(t_{i}\right)\right)=\alpha EWMA\left(Z\left(t_{i-1}\right)\right)+\left(1-\alpha\right)Z\left(t_{i}\right)\label{EQ_ewma}
\end{equation}
where $EWMA\left(Z\left(t_{1}\right)\right)=Z\left(t_{1}\right)$,
$i=2,3,4...$. Here, $t_{1}$ is the time of the first update in the
variable $Z$ at the start of each day. EWMA's are computed for each
day. With abuse of notation we then set the raw covariate equal to
$EWMA\left(Z\left(t_{i}\right)\right)$. The raw covariates are similar
to the ones in Mucciante and Sancetta (2022), though with some differences.
Hence, we briefly summarize their construction. The book volume imbalance
at level $j$ is defined as 
\begin{equation}
{\rm VolImb}_{j}=\frac{{\rm BidSize}_{j}-{\rm AskSize}_{j}}{{\rm BidSize}_{j}+{\rm AskSize}_{j}}\label{EQ_VolImb}
\end{equation}
where ${\rm BidSize}_{j}$ is the bid size (quantity) at level $j$,
and similarly for ${\rm AskSize}_{j}$. This variable takes values
in $\left[-1,1\right]$. The trade imbalance is computed from the
EWMA of the signed traded volume every time there is a trade. We then
divide it by the EWMA of the unsigned volumes. The EWMA's parameter
is $\alpha=0.98$ for both denominator and numerator. Durations are
in seconds with nanosecond decimals. They are then passed to EWMA
filters with parameters $\alpha=0.98$ and $0.90$. The spread is
computed in basis points. After the application of EWMA's filters,
our model (\ref{EQ_intensityRepresentation}) has 7 raw covariates
per instrument plus a seasonal component. The seasonal is time of
the day for each update between $\left[09:30,16:30\right]$ EST, standardized
to be in $\left[0,1\right]$. We also include information from two
additional auxiliary instruments with no seasonal component. Then,
the total number of raw covariates is 22. We then apply one hot encoding
as discussed in Section \ref{Section_oneHotEncodingEmpiricalDetails}.
The total number of parameters to estimate becomes at most 168 once
we add a constant. Adding a constant introduces perfect linear dependence.
As discussed at the end of Section \ref{Section_estimation} this
is not a problem from a computational point of view. 

We define a reference time to be the time at which a book or trade
update is sent for a traded instrument. A traded instrument is the
one whose intensity we are modelling. Then, to limit the size of the
data, after computing all the covariates, we sample them at the reference
time only. Finally, to ensure that the covariates are predictable,
we make them left continuous by lagging them as a very last step in
the procedure. Failing to do so would lead to forward looking bias.
To see this, let $t_{i-1}$ and $t_{i}$ be the time of the $i-1$
and $i$ order book update. The intensity at time $t_{i}$ can only
use the book update from time $t_{i-1}$. It will use the information
from the $i^{th}$ book update immediately after $t_{i}$. This implies
that if we were to observe a trade at time $t_{i}$, the $i^{th}$
book update could not be used to predict the trade. In summary, we
ensure that the same conditions for live trading are reproduced in
our estimation. 

\begin{table}[H]
\caption{Raw Covariates Used for Estimation. The column ``Smoothing'' reports
the smoothing parameter used if an EWMA had been applied to the original
variable. }
\label{Table_covariatesEmpiricalSection}

\begin{tabular}{llllc}
 &  &  &  & \tabularnewline
Variables & \multicolumn{2}{c}{Short Name} & Smoothing & \tabularnewline
\cline{1-4}
Seasonal & Seas &  &  & \tabularnewline
Volume Imbalance Level 1 & VolImb1 &  &  & \tabularnewline
Volume Imbalance Level 2 & VolImb2 &  &  & \tabularnewline
Volume Imbalance Level 3 & VolImb3 &  &  & \tabularnewline
Spread & Spread &  &  & \tabularnewline
Trade Imbalance & TrdImb98 &  & $\alpha=0.98$ & \tabularnewline
Durations & Dur98, & Dur90 & $\alpha=0.98$, $0.90$ & \tabularnewline
 &  &  &  & \tabularnewline
\end{tabular}
\end{table}

\subsection{One-Hot Encoding of Covariates \label{Section_oneHotEncodingEmpiricalDetails}}

To automate the procedure, we opted for a simple rule that can be
applied to all the covariates. Let $q_{x}$ be the $x\%$ quantile
of a covariate based on the estimation sample. For all covariates,
we consider the following bins: $\left[-\infty,q_{1}\right)$, $\left[q_{1},q_{10}\right)$,
$\left[q_{10},q_{25}\right)$, $\left[q_{25},q_{50}\right)$, $\left[q_{50},q_{75}\right)$,
$\left[q_{75},q_{90}\right)$, $\left[q_{90},q_{99}\right)$, $\left[q_{99},\infty\right)$.
If for each covariate the quantiles are are not unique, we take the
set of unique quantile and construct bins accordingly. For example
suppose that for the spread $q_{50}=q_{75}=q_{90}$ while all other
quantiles are unique. Then, the bins we use for the spread are $\left[-\infty,q_{1}\right)$,
$\left[q_{1},q_{10}\right)$, $\left[q_{10},q_{25}\right)$, $\left[q_{25},q_{50}\right)$,
$\left[q_{50},\infty\right)$. Also note that binning in $\left[-\infty,q_{1}\right)$
or $\left[0,q_{1}\right)$ produces the same result for covariates
that take nonnegative values. According to the aforementioned binning
strategy, the total number of parameters to be estimated, including
a parameter for a constant, is at most 177 when all quantiles are
unique. However, due to nonuniqueness of some of the quantiles for
the spread, the actual number of parameters to estimate becomes 168. 

\subsection{The Models\label{Section_models}}

To understand the importance of the nonlinearity in the impact of
order book variables and the self exciting nature of the trading events,
we consider a number of model specifications within the current framework.
We list these next.\\
\\
E:	$h\left(t\right)=1$, $g\left(t\right)=X\left(t\right)'b$ (no
self excitation, $Z\mapsto X$ by one-hot encoding)\\
H01: $h\left(t\right)$ as in (\ref{EQ_hawkes_ht}) with $L=1$, $g\left(t\right)=1$
(no book information)\\
H02:	$h\left(t\right)$ as in (\ref{EQ_hawkes_ht}) with $L=2$, $g\left(t\right)=1$
(no book information)\\
H1:	$h\left(t\right)$ as in (\ref{EQ_hawkes_ht}) with $L=1$, $g\left(t\right)=X\left(t\right)'b$
($Z\mapsto X$ by one-hot encoding)\\
H2:	$h\left(t\right)$ as in (\ref{EQ_hawkes_ht}) with $L=2$, $g\left(t\right)=X\left(t\right)'b$
($Z\mapsto X$ by one-hot encoding)\\
H1L:	$h\left(t\right)$ as in (\ref{EQ_hawkes_ht}) with $L=1$, $g\left(t\right)=Z\left(t\right)'b$
(linear raw covariates with $b$ in the reals)\\
H2L:	$h\left(t\right)$ as in (\ref{EQ_hawkes_ht}) with $L=2$, $g\left(t\right)=Z\left(t\right)'b$
(linear raw covariates with $b$ in the reals)\\

In the above, the raw covariates in Table \ref{Table_covariatesEmpiricalSection}
are denoted by $Z$. For the nonlinear models, these are mapped to
$X$ via one-hot encoding. The model is then estimated under a positivity
constraint and under an upper bound constraint. In particular, model
E corresponds to no self excitation, so that conditioning on $g\left(t\right)$
the hazard functions is the one of an exponential distribution. Models
H01 and H02 are Hawkes processes with kernel equal to the sum of one
and two exponential functions, respectively. Models H1 and H2 are
like H01 and H02, respectively, times $g\left(t\right)$, i.e. they
include order book information. Models H1L and H2L are like H1 and
H2, respectively, but only allow linear impact of the raw covariates.
In this case the $b$ coefficient is allowed to be negative so that
the impact can have a negative sign. Then, to avoid a negative intensity,
we impose a floor on the intensity when testing out of sample. Additional
details on estimation constraints can be found in Section \ref{Section_paramsRestrictionsEmpirical}
of the Appendix.

\subsection{Results\label{Section_results}}

Data summary statistics show that the size of the data is large across
all instruments, though there is some degree of variation (Table \ref{Table_summaryStats}).
For example, during liquid periods, trades information is disseminated
even every few microseconds. 

We estimate our models for buy and sell events separately. The positivity
constraint and the constraint on the sum of the coefficients lead
to a relatively sparse estimator. For the model with $L=1$ in (\ref{EQ_hawkes_ht}),
which we defined as H1 in Section \ref{Section_models}, the average
number of nonzero estimated coefficients is roughly between 20\% to
30\% across the four stocks both for Any Trade Arrivals and the Large
Trade Arrivals. 

We inspect the number of nonzero coefficient for each covariate. Covariates
for which the coefficients of the one-hot encoding are all zero do
not enter the model and could be deemed as unimportant. We note that
there is a reasonable level of consistency across the different stocks.
The detailed results are reported in Tables \ref{Table_nActiveWithNameAnyTrades}
and \ref{Table_nActiveWithNameLargeTrades}.

We conduct the test of Section \ref{Section_ComparingIntensities}
in order to answer the questions from Section \ref{Section_empirical}.
For example, we write E-H1 to mean that that the loglikelihood ratio
is constructed as the likelihood of model E minus the loglikelihood
of model H1 (see Section \ref{Section_models}). Then, we compute
the following test statistics:
\begin{enumerate}
\item H01-H1, H02-H1, and H02-H2: A large negative value of the test statistic
implies that the order book provide information in addition to what
is captured by the self exciting nature of event arrivals;
\item E-H1: a large negative value means that self excitation is important
even after accounting for order book information;
\item H1L-H1, H2L-H1, H1L-H2, and H2L-H2: A large negative value means that
the impact of the order book is nonlinear irrespective of the kernel
chosen in (\ref{EQ_hawkes_ht});
\item H2-H1: A large positive value means that the simple exponential kernel
is not sufficient to capture the self excitation of the trading events. 
\end{enumerate}
We find that accounting for both self excitation and order book information
(models H1 and H2) is important (Points 1 and 2). We also find that
the impact of of the order book is nonlinear (Point 3). Finally, the
self exciting nature of trading events usually requires the use of
a kernel more complex that the simple exponential one (Point 4). The
details from the tests are in Table \ref{Table_testRestriction}.
The values of the test statistics are very large in absolute value
because the sample size $T$ is large (see the remarks about the power
of the test just before Theorem \ref{Theorem_testCompetingIntensities}). 

\begin{figure}
\caption{CSCO Volume Imbalance Level 1. The model is for $L=1$ in (\ref{EQ_hawkes_ht})
for CISCO and Any Trade Arrivals of buy trades. The estimated coefficients
$b_{k}$ for VolImb1 are plotted on the Y-axis as a function of $k$,
which is the bin number out of 8 bins based on quantiles (see Section
\ref{Section_oneHotEncodingEmpiricalDetails}). Bin numbers greater
than 4 correspond to positive values of ${\rm VolImb1}$.}
\label{Figure_CLVolImb1_sample1}

\includegraphics[scale=0.7]{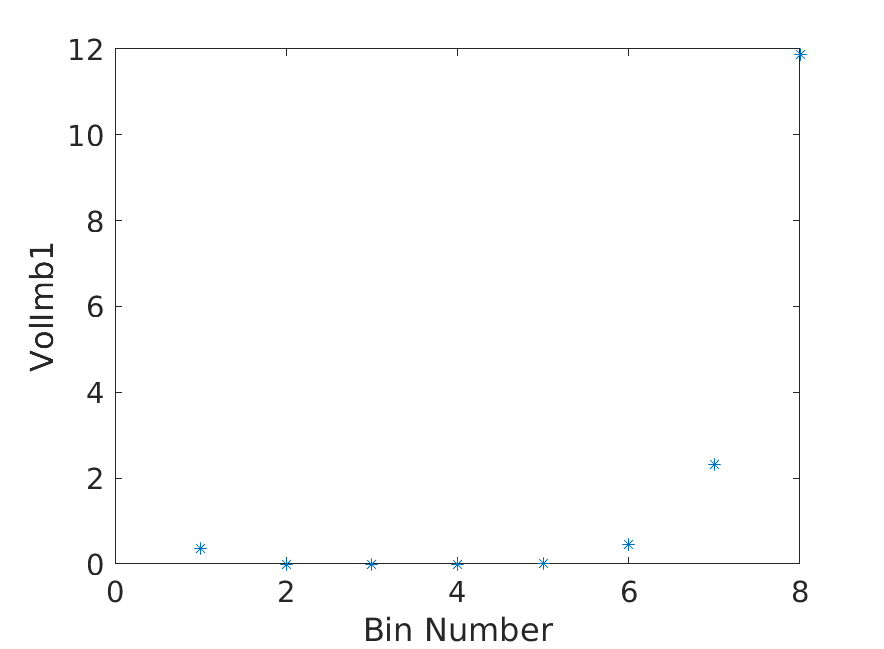}
\end{figure}

\begin{figure}
\caption{CSCO Duration (Dur98). The model is for $L=1$ in (\ref{EQ_hawkes_ht})
for CISCO and Any Trade Arrivals of buy trades. The estimated coefficients
$b_{k}$ for Dur98 are plotted on the Y-axis as a function of $k$,
which is the bin number out of 8 bins based on quantiles (see Section
\ref{Section_oneHotEncodingEmpiricalDetails}). For example, bin number
1 corresponds to ${\rm Dur98}$ smaller that the 1\% quantile.}
\label{Figure_CLDur98}

\noindent\includegraphics[scale=0.7]{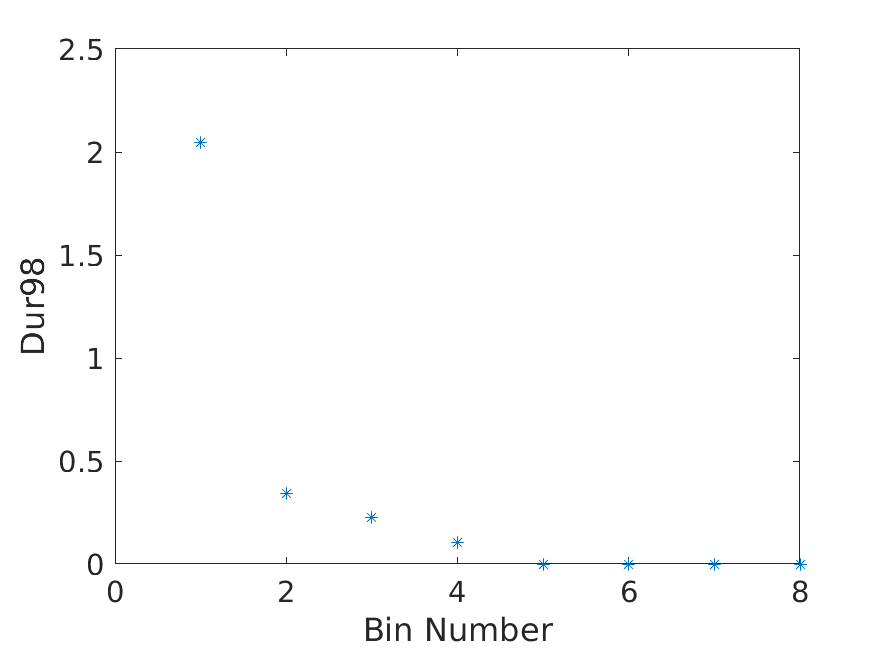}
\end{figure}

We conclude mentioning that one-hot encoding can produce plots that
are interpretable and regular even without any constraint (Figures
\ref{Figure_CLVolImb1_sample1} and \ref{Figure_CLDur98}). 

\begin{table}
\caption{Sample Size Statistics. The total number N of events that correspond
to Any Trade Arrivals and Large Trade Arrivals is reported together
with the total number m of book updates with unique time stamps. The
number of days is 42 for the period 01/Mar/2019 - 30/04/2019. }
\label{Table_summaryStats}

\begin{tabular}{cccc}
 & \multicolumn{2}{c}{N} & m\tabularnewline
 & Any Trade Arrivals & Large Trade Arrivals & \tabularnewline
\cline{2-4}
AMZN & 631,370 & 407,130 & 17,130,000\tabularnewline
CSCO & 295,220 & 107,930 & 27,943,000\tabularnewline
DIS & 493,100 & 292,820 & 24,938,000\tabularnewline
KO & 121,210 & 52842 & 13,956,000\tabularnewline
 &  &  & \tabularnewline
\end{tabular}
\end{table}

\begin{table}
\begin{raggedright}
\caption{Active Covariates for Any Trade Arrivals. The total number of estimated
non zero coefficients (from one-hot encoding) for each covariate (as
in Table \ref{Table_covariatesEmpiricalSection}) are reported. The
total sum (Sum) of nonzero coefficients and their relative proportion
(Proportion), out of the estimated 168 parameters, are also reported.
The models considered are E, H1 and H2, as defined in Section \ref{Section_models}.}
\label{Table_nActiveWithNameAnyTrades}
\par\end{raggedright}
\resizebox{1\textwidth}{!}{

\begin{tabular}{lcccccclcccccc}
 &  &  &  &  &  &  &  &  &  &  &  &  & \tabularnewline
 & \multicolumn{3}{c}{AMZN} & \multicolumn{3}{c}{CSCO} &  & \multicolumn{3}{c}{DIS} & \multicolumn{3}{c}{KO}\tabularnewline
 & E & H1 & H2 & E & H1 & H2 &  & E & H1 & H2 & E & H1 & H2\tabularnewline
\hline 
AMZN\_Dur90 & 2 & 3 & 3 & 2 & 2 & 2 & DIS\_Dur90 & 3 & 4 & 4 & 1 & 3 & 2\tabularnewline
AMZN\_Dur98 & 3 & 3 & 3 & 3 & 3 & 3 & DIS\_Dur98 & 4 & 4 & 4 & 1 & 3 & 3\tabularnewline
AMZN\_Seas & 4 & 5 & 5 & 1 & 2 & 2 & DIS\_Seas & 2 & 3 & 3 & 0 & 1 & 1\tabularnewline
AMZN\_Spread & 4 & 4 & 4 & 0 & 1 & 1 & DIS\_Spread & 1 & 1 & 1 & 1 & 1 & 1\tabularnewline
AMZN\_TrdImb98 & 3 & 3 & 3 & 0 & 0 & 0 & DIS\_TrdImb98 & 3 & 2 & 2 & 1 & 0 & 0\tabularnewline
AMZN\_VolImb1 & 0 & 0 & 0 & 0 & 0 & 0 & DIS\_VolImb1 & 4 & 4 & 4 & 3 & 3 & 3\tabularnewline
AMZN\_VolImb2 & 0 & 0 & 0 & 0 & 0 & 0 & DIS\_VolImb2 & 3 & 3 & 3 & 1 & 1 & 1\tabularnewline
AMZN\_VolImb3 & 0 & 0 & 0 & 3 & 3 & 3 & DIS\_VolImb3 & 2 & 2 & 2 & 3 & 4 & 4\tabularnewline
CSCO\_Dur90 & 1 & 1 & 1 & 4 & 4 & 4 & KO\_Dur90 & 0 & 0 & 0 & 4 & 4 & 4\tabularnewline
CSCO\_Dur98 & 3 & 3 & 3 & 5 & 5 & 5 & KO\_Dur98 & 1 & 1 & 1 & 3 & 4 & 4\tabularnewline
CSCO\_Spread & 1 & 1 & 1 & 0 & 0 & 0 & KO\_Spread & 0 & 0 & 0 & 1 & 1 & 1\tabularnewline
CSCO\_TrdImb98 & 1 & 0 & 0 & 2 & 2 & 2 & KO\_TrdImb98 & 0 & 0 & 0 & 2 & 2 & 2\tabularnewline
CSCO\_VolImb1 & 3 & 3 & 3 & 4 & 5 & 5 & KO\_VolImb1 & 1 & 1 & 1 & 5 & 5 & 5\tabularnewline
CSCO\_VolImb2 & 2 & 3 & 3 & 3 & 3 & 3 & KO\_VolImb2 & 0 & 0 & 0 & 3 & 3 & 3\tabularnewline
CSCO\_VolImb3 & 0 & 0 & 0 & 1 & 2 & 2 & KO\_VolImb3 & 0 & 0 & 0 & 1 & 1 & 1\tabularnewline
SPY\_Dur90 & 3 & 3 & 3 & 3 & 3 & 3 & SPY\_Dur90 & 2 & 2 & 2 & 4 & 4 & 4\tabularnewline
SPY\_Dur98 & 4 & 4 & 4 & 2 & 3 & 3 & SPY\_Dur98 & 1 & 1 & 1 & 3 & 3 & 3\tabularnewline
SPY\_Spread & 2 & 2 & 2 & 2 & 2 & 2 & SPY\_Spread & 2 & 2 & 2 & 2 & 2 & 2\tabularnewline
SPY\_TrdImb98 & 1 & 2 & 2 & 1 & 1 & 1 & SPY\_TrdImb98 & 1 & 1 & 1 & 2 & 2 & 2\tabularnewline
SPY\_VolImb1 & 3 & 3 & 3 & 2 & 3 & 3 & SPY\_VolImb1 & 1 & 3 & 3 & 3 & 3 & 3\tabularnewline
SPY\_VolImb2 & 1 & 1 & 1 & 1 & 1 & 1 & SPY\_VolImb2 & 0 & 0 & 0 & 1 & 1 & 1\tabularnewline
SPY\_VolImb3 & 0 & 1 & 1 & 1 & 1 & 1 & SPY\_VolImb3 & 0 & 0 & 0 & 2 & 1 & 1\tabularnewline
Sum & 41 & 45 & 45 & 40 & 46 & 46 &  & 31 & 34 & 34 & 47 & 52 & 51\tabularnewline
Proportion & 0.24 & 0.27 & 0.27 & 0.24 & 0.27 & 0.27 &  & 0.18 & 0.20 & 0.20 & 0.28 & 0.31 & 0.30\tabularnewline
\end{tabular}

}
\end{table}

\begin{table}
\begin{raggedright}
\caption{Active Covariates for Large Trade Arrivals. The total number of estimated
non zero coefficients (from one-hot encoding) for each covariate (as
in Table \ref{Table_covariatesEmpiricalSection}) are reported. The
total sum (Sum) of nonzero coefficients and their relative proportion
(Proportion), out of the estimated 168 parameters, are also reported.
The models considered are E, H1 and H2, as defined in Section \ref{Section_models}.}
\label{Table_nActiveWithNameLargeTrades}
\par\end{raggedright}
\resizebox{1\textwidth}{!}{

\begin{tabular}{lcccccclcccccc}
 &  &  &  &  &  &  &  &  &  &  &  &  & \tabularnewline
 & \multicolumn{3}{c}{AMZN} & \multicolumn{3}{c}{CSCO} &  & \multicolumn{3}{c}{DIS} & \multicolumn{3}{c}{KO}\tabularnewline
 & E & H1 & H2 & E & H1 & H2 &  & E & H1 & H2 & E & H1 & H2\tabularnewline
\hline 
AMZN\_Dur90 & 2 & 2 & 2 & 2 & 2 & 2 & DIS\_Dur90 & 3 & 3 & 3 & 1 & 1 & 1\tabularnewline
AMZN\_Dur98 & 3 & 3 & 3 & 2 & 2 & 2 & DIS\_Dur98 & 3 & 3 & 3 & 2 & 2 & 2\tabularnewline
AMZN\_Seas & 4 & 4 & 4 & 3 & 3 & 3 & DIS\_Seas & 4 & 3 & 3 & 2 & 2 & 2\tabularnewline
AMZN\_Spread & 4 & 4 & 4 & 1 & 0 & 0 & DIS\_Spread & 1 & 1 & 1 & 1 & 1 & 1\tabularnewline
AMZN\_TrdImb98 & 2 & 2 & 2 & 0 & 0 & 0 & DIS\_TrdImb98 & 2 & 2 & 2 & 0 & 0 & 0\tabularnewline
AMZN\_VolImb1 & 4 & 4 & 4 & 0 & 0 & 0 & DIS\_VolImb1 & 4 & 4 & 4 & 1 & 1 & 1\tabularnewline
AMZN\_VolImb2 & 0 & 0 & 0 & 0 & 0 & 0 & DIS\_VolImb2 & 2 & 3 & 3 & 1 & 1 & 1\tabularnewline
AMZN\_VolImb3 & 0 & 0 & 0 & 2 & 1 & 1 & DIS\_VolImb3 & 1 & 1 & 1 & 2 & 2 & 2\tabularnewline
CSCO\_Dur90 & 0 & 1 & 1 & 4 & 4 & 4 & KO\_Dur90 & 0 & 0 & 0 & 4 & 3 & 3\tabularnewline
CSCO\_Dur98 & 3 & 3 & 3 & 3 & 3 & 3 & KO\_Dur98 & 1 & 1 & 1 & 3 & 3 & 4\tabularnewline
CSCO\_Spread & 1 & 1 & 1 & 0 & 0 & 0 & KO\_Spread & 0 & 0 & 0 & 0 & 0 & 0\tabularnewline
CSCO\_TrdImb98 & 0 & 0 & 0 & 1 & 0 & 0 & KO\_TrdImb98 & 0 & 0 & 0 & 1 & 0 & 0\tabularnewline
CSCO\_VolImb1 & 2 & 2 & 2 & 3 & 3 & 3 & KO\_VolImb1 & 0 & 1 & 1 & 3 & 3 & 3\tabularnewline
CSCO\_VolImb2 & 2 & 2 & 2 & 2 & 2 & 2 & KO\_VolImb2 & 0 & 0 & 0 & 1 & 0 & 0\tabularnewline
CSCO\_VolImb3 & 0 & 0 & 0 & 1 & 1 & 1 & KO\_VolImb3 & 0 & 0 & 0 & 1 & 0 & 0\tabularnewline
SPY\_Dur90 & 3 & 3 & 3 & 3 & 3 & 3 & SPY\_Dur90 & 2 & 2 & 2 & 3 & 3 & 3\tabularnewline
SPY\_Dur98 & 4 & 4 & 4 & 3 & 3 & 3 & SPY\_Dur98 & 1 & 1 & 1 & 3 & 3 & 3\tabularnewline
SPY\_Spread & 2 & 2 & 2 & 2 & 2 & 2 & SPY\_Spread & 2 & 2 & 2 & 2 & 2 & 2\tabularnewline
SPY\_TrdImb98 & 1 & 1 & 1 & 1 & 1 & 1 & SPY\_TrdImb98 & 1 & 1 & 1 & 2 & 2 & 2\tabularnewline
SPY\_VolImb1 & 3 & 3 & 3 & 3 & 3 & 3 & SPY\_VolImb1 & 1 & 2 & 2 & 3 & 3 & 3\tabularnewline
SPY\_VolImb2 & 1 & 1 & 1 & 1 & 1 & 1 & SPY\_VolImb2 & 0 & 0 & 0 & 1 & 0 & 0\tabularnewline
SPY\_VolImb3 & 1 & 1 & 1 & 1 & 0 & 0 & SPY\_VolImb3 & 0 & 0 & 0 & 0 & 0 & 0\tabularnewline
Sum & 42 & 43 & 43 & 38 & 34 & 34 &  & 28 & 30 & 30 & 37 & 32 & 33\tabularnewline
Proportion & 0.25 & 0.26 & 0.26 & 0.23 & 0.20 & 0.20 &  & 0.17 & 0.18 & 0.18 & 0.22 & 0.19 & 0.20\tabularnewline
\end{tabular}

}
\end{table}

\begin{table}
\caption{Test of Model Restrictions. The values for the test statistic in (\ref{EQ_testRejectRule})
are reported. The test is constructed so that the last 5 trading days
are used as test and the previous days for estimation. Models are
defined in Section \ref{Section_models}. The null hypothesis is that
two models perform the same out of sample. A large positive value
means that the null is rejected and the first model performs better
than the second. The reverse applies for large negative values. The
statistic is standard normal, and the ``simple'' models are almost
always rejected. }
\label{Table_testRestriction}

\begin{tabular}{lccccccccc}
 &  &  &  &  &  &  &  &  & \tabularnewline
 &  & \multicolumn{4}{c}{Any Trade Size} & \multicolumn{4}{c}{Large Trade Size}\tabularnewline
 &  & AMZN & CSCO & DIS & KO & AMZN & CSCO & DIS & KO\tabularnewline
\hline 
E-H1 & Buy & -109.88 & -70.20 & -93.44 & -44.41 & -81.40 & -22.48 & -66.48 & -16.57\tabularnewline
 & Sell & -112.77 & -72.18 & -86.61 & -43.36 & -82.20 & -26.36 & -59.26 & -20.23\tabularnewline
H01-H1 & Buy & -20.28 & -7.84 & -6.43 & -12.83 & -37.01 & -35.13 & -14.08 & -21.27\tabularnewline
 & Sell & -15.53 & -8.32 & -9.65 & -39.54 & -35.27 & -41.36 & -21.21 & -38.94\tabularnewline
H02-H1 & Buy & -20.96 & -3.10 & 39.23 & -4.11 & -20.53 & -33.75 & -4.24 & -21.34\tabularnewline
 & Sell & -15.28 & -6.82 & 35.15 & -24.35 & -18.58 & -39.07 & -12.37 & -38.00\tabularnewline
H2-H1 & Buy & -73.86 & -44.02 & 63.92 & 27.65 & 41.51 & 11.13 & 20.55 & 8.91\tabularnewline
 & Sell & -54.83 & 49.30 & 59.82 & 18.56 & 42.20 & 10.39 & 27.04 & 7.38\tabularnewline
H1L-H1 & Buy & -27.75 & -13.95 & -7.61 & -11.54 & -27.35 & -25.92 & -10.68 & -23.18\tabularnewline
 & Sell & -24.68 & -15.76 & -2.95 & -39.38 & -26.78 & -32.51 & -8.68 & -41.81\tabularnewline
H2L-H1 & Buy & -27.99 & -12.27 & 29.98 & 2.42 & -20.67 & -25.14 & -5.88 & -23.05\tabularnewline
 & Sell & -24.42 & -17.14 & 33.22 & -22.65 & -19.54 & -31.02 & -3.14 & -40.92\tabularnewline
H02-H2 & Buy & -19.02 & -2.77 & -1.07 & -18.27 & -38.17 & -34.68 & -14.76 & -22.12\tabularnewline
 & Sell & -15.09 & -9.02 & -5.28 & -34.93 & -36.76 & -40.34 & -26.29 & -40.28\tabularnewline
H1L-H2 & Buy & -26.70 & -13.71 & -40.41 & -24.05 & -34.81 & -26.44 & -16.76 & -23.96\tabularnewline
 & Sell & -24.57 & -17.45 & -38.13 & -40.26 & -34.61 & -33.22 & -16.58 & -42.91\tabularnewline
H2L-H2 & Buy & -26.95 & -12.03 & -3.63 & -10.28 & -29.10 & -25.79 & -12.24 & -23.88\tabularnewline
 & Sell & -24.31 & -18.82 & -0.90 & -32.08 & -28.33 & -32.07 & -11.16 & -43.16\tabularnewline
 &  &  &  &  &  &  &  &  & \tabularnewline
\end{tabular}
\end{table}

\section{Conclusion\label{Section_conclusion}}

This paper presented a Hawkes process augmented by order book information.
The model is the standard Hawkes process with a multiplicative term
which is a function of order book variables. We showed conditions
under which the process is stationary and ergodic. We focused on a
flexible estimation procedure suitable for large datasets where the
results are however intuitively simple to interpret. Using theoretical
results as well as simulation examples we were able to justify the
procedure. The results showed that the convergence rates only deteriorate
at a logarithmic rate with the number of parameters. The motivation
for the algorithm presented here and its study is to allow us to use
flexible techniques to model the impact of the covariates derived
from the order book, possibly using very large datasets. In particular,
we focus on one-hot encoding. This maps the original covariates into
a higher dimensional space, a case covered by our methodology. Our
application to four stocks traded on the NYSE showed the importance
of using order book information. In summary, our sample testing procedure
shows that nonlinearity of the order book adds value to the self exciting
nature of high frequency event arrivals. 

\newpage{}

\part*{Appendix}

\setcounter{figure}{0} \renewcommand{\thefigure}{A.\arabic{figure}}

\setcounter{equation}{0} \renewcommand{\theequation}{A.\arabic{equation}}

\setcounter{section}{0} \renewcommand{\thesection}{A.\arabic{section}}

In Section \ref{Section_exogenousEndogenousTrading}, we give further
information on the interpretation of the model in terms of exogenous
and endogenous information arrival. Section \ref{Section_proofs}
proves all the results stated in the paper. Section \ref{Section_AppendixAdditionalDetails}
includes further details regarding the empirical application. 

\section{Model Interpretation: Exogenous Information Versus Endogenous Market
Activity\label{Section_exogenousEndogenousTrading}}

We expand on the model representation in terms of an immigration birth
process (Hawkes and Oakes, 1974). In our context, immigration means
exogenous information arrival, while birth represents endogenously
generated information. When $g_{0}$ is constant, exogenous information
is information that is independent of arrivals, i.e. independent of
$N$, hence strictly exogenous. Endogenous information at time $t$
is the information that has been generated by the trading activity
up to time $t$, i.e. $\left(N\left(s\right)\right))s<t$. Here this
interpretation is generalized. 

Write $\kappa\left(t\right):=\sum_{l=1}^{L}d_{0,l}e^{-a_{0,l}t}$.
This is the kernel of the process in (\ref{EQ_hawkes_ht}). Let $T_{n}$
be the $n^{th}$ event arrival, and suppose that this is the latest
arrival time before time $t$. Then, $g_{0}\left(t\right)c_{0}$ represents
the unit rate of arrival of immigrants at time $t$. Each immigrant
has descendants who procreate ad infinitum. If $g_{0}$ were constant,
the immigration arrival would be strictly exogenous, as in the standard
Hawkes process. Here, the intensity of immigrants arrivals depends
on what we call the environment, and it is captured by the process
$g_{0}$ at time $t$. 

At time $t$ there are $n$ independent individuals in the population.
Nobody dies, but they all age. At time $t$, the $i^{th}$ individual
has age $t-T_{i}$ and the unit intensity at which they give birth
is $g_{0}\left(t\right)\kappa\left(t-T_{i}\right)$. Given that $\kappa$
is decreasing, an older individual has lower probability of giving
birth. However, their chances can be increased if $g_{0}\left(t\right)$
is higher than normal. Hence, the introduction of the process $g_{0}$
makes the individuals in the populations dependent, unlike the standard
Hawkes process. Conditional on the environment, i.e. conditioning
on $g_{0}\left(t\right)$, the probability of giving birth is independent
across the individuals in the population. Conditioning on $g_{0}\left(t\right)$,
the intensity of new arrivals in the the population, either via immigration
or birth is $g_{0}\left(t\right)c_{0}+g_{0}\left(t\right)\sum_{i=1}^{n}\kappa\left(t-T_{i}\right)$.
This is exactly (\ref{EQ_intensityRepresentation}) once we substitute
the definition of $\kappa$. 

Suppose that $T_{0}$ is the time of arrival of an immigrant. During
their infinite lifetime, their expected total number of offspring
is $n_{g_{0}}:=\mathbb{E}_{T_{0}}\int_{T_{0}}^{\infty}g\left(t\right)\kappa\left(t-T_{0}\right)dt$,
where $\mathbb{E}_{T_{0}}$ is expectation conditional on the information
at time $T_{0}$. If $g_{0}$ is constant, $n_{g_{0}}$ does not depend
on $T_{0}$. Then, the number $n_{g_{0}}$ is called branching ratio.
When less than one, the probability of extinction is one. This is
necessary to ensure that the effect of any event arrival eventually
dies out so that the process is stationary. If $g_{0}$ is not constant,
but uniformly bounded, we still have $n_{g_{0}}<1$ if $\sup_{t}\left|g_{0}\left(t\right)\right|_{\infty}\left(\sum_{l=1}^{L}d_{0,l}/a_{0,l}\right)^{-1}<1$,
where $\sum_{l=1}^{L}d_{0,l}/a_{0,l}=\int_{0}^{\infty}\kappa\left(t\right)dt$. 

In the context of the present paper, exogenous news are represented
by the immigrants. However, information is costly and the goal of
an informed participant is to trade reveling as little information
as possible (Grossman and Stiglitz, 1980). Trading is also costly,
so that the trading strategy of both informed and uninformed traders
must depend on the environment, e.g. the state of the order book.
Optimal execution strategies used to minimize private information
transmission and trading cost needs to take into account the order
book and market microstructures. In a frictionless market where cost
of information is zero, prices would adjusts immediately and not result
in any prolonged trading activity. In the context of a Hawkes process,
this means that each immigrant would have no descendants and trading
is only kept alive by exogenous news events. For examples, this is
the case when $\min_{l}a_{0,l}\rightarrow\infty$ in (\ref{EQ_hawkes_ht}).
On the opposite side of the spectrum, information is never fully absorbed
by the market and trading is kept alive by the descendants and an
equilibrium price is never reached. Clearly, this is unnatural. Eventually
information is absorbed by the market and trading is only kept alive
by exogenous information transmission. This means that $n_{g_{0}}<1$.
The extent to which $n_{g_{0}}$ is close to one tells us about the
level of endogenous market interaction that spawned out of any news.
Equivalently, it represents the average fraction of endogenously generated
events. Clearly a market where information is not quickly absorbed
and trading is just the result of endogenous interaction (numerous
descendants from each immigrant) is a market prone to fragility. Filomonov
and Sornette (2015) call this market reflexivity.

\section{Proofs\label{Section_proofs}}

The intensity $\lambda_{0}\left(t\right)=\lambda_{0}\left(\omega,t\right)$
is a continuous time stochastic process, i.e. a function of two variables
$t\geq0$ and $\omega\in\Omega$ where $\left(\Omega,\mathcal{B},P\right)$
is a probability space and for each $t$, $\lambda_{0}\left(t,\cdot\right)$
is measurable on $\Omega$. Similarly $h_{0}\left(t\right)=h_{0}\left(\omega,t\right)$
and $X\left(t\right)=X\left(\omega,t\right)$. The covariate process
$X$ and the baseline intensity are predictable processes. These quantities
are supposed to be left continuous with right hand limits. For ease
of notation, we may freely switch between $\lambda_{0}\left(t\right)$
and $h_{0}\left(t\right)g_{0}\left(t\right)$ and compactly write
$\lambda_{0}=h_{0}g_{0}$. To make the notation simpler and more readable,
we shall write $Ph_{0}g_{0}$ to mean $\int_{\Omega}h_{0}\left(\omega,0\right)g_{0}\left(\omega,0\right)dP\left(\omega\right)$
and similarly for other quantities. 

Given that all the quantities are finite dimensional, both $Q_{T}$
and $R_{T}$ (in (\ref{EQ_quadraticRisk}) and (\ref{EQ_contrastFunctionUnfeasible-1}))
are Frech\'{e}t differentiable. Throughout to ease notation, for
any function $f$ that is bounded below by a constant, we use $c_{f}$
to indicate that constant. If bounded above by a constant, we use
$C_{f}$ to indicate such constant. If $\mathcal{F}$ is a class of
functions bounded below and/or above by constants, we denote such
constants by $c_{\mathcal{F}}$ and $C_{\mathcal{F}}$ respectively.
Finally, to avoid making reference to constants, we use $\lesssim$
when the left hand side (l.h.s.) is bounded by a constant times the
right hand side (r.h.s.) and $\asymp$ when the l.h.s. is bounded
below and above by constants times the r.h.s.. 

We prove Theorem \ref{Theorem_stationarityErgodicity} first, followed
by Theorem \ref{Theorem_convergence} and then \ref{Theorem_QuadraticLossConsistency},
as the latter easily follows from the former. We then prove Theorem
\ref{Theorem_testCompetingIntensities}.

\subsection{Proof of Theorem \ref{Theorem_stationarityErgodicity}}

We shall use as much as possible the notation in Br\'{e}maud and
Massouli\'{e} (1996), BM96 henceforward. Throughout, we view $N$
as an element in $\mathbb{M}$, the space of Radon measures on $\mathbb{R}$.
Let $\omega$ be a Poisson point process associated with the arrival
times $\left(\tau_{i}\right)_{i\geq1}$. Theorem 4 and Lemma 3 in
BM96 show that, iteratively for $n=0,1,2...$, given $\omega$ and
an intensity $\lambda^{n}$ we can construct a point process $N^{n}$
that has such intensity (BM96, eq. 14). We use the same notation as
in BM96, hence here $N^{n}$ and $\lambda^{n}$ are sequences and
should not be confused with $n$ powers. The intensity is constructed
to satisfy $\lambda^{n+1}\left(t\right)=\phi\left(\int_{\left(-\infty,t\right)}\sum_{l=1}^{L}d_{0,l}e^{-a_{0,l}\left(t-s\right)}dN^{n}\left(s\right)\right)$
for $\phi:\mathbb{R}\rightarrow\left[0,\infty\right)$ and such that
$\phi\left(x\right)\leq\alpha+\beta x$ for some positive constants
$\alpha,\beta$. In our case $\phi$ is replaced by a predictable
process $\phi_{t}^{n}\left(x\right)=g_{0}^{n}\left(t\right)\left(c_{0}+x\right)$
where, by the Regularity Conditions, $g_{0}^{n}\left(t\right)=f_{0}\left(W\left(t\right),S_{t}N^{n}\right)$.
It also follows that because $\omega$ is stationary for the shift
operator, so is $N^{n}$ (BM96, first and last paragraphs on p.1273
and 1274, respectively). Hence, $g_{0}^{n}\left(t\right)$ is stationary
for every iteration $n$ by Condition \ref{Condition_endogeneity}.
Using the upper bound for $\phi_{t}^{n}\left(\cdot\right)$ and stationarity
of $\lambda^{n}$, we have that, for $t=0$, 
\begin{align*}
\mathbb{E}\lambda^{n+1}\left(0\right)= & \mathbb{E}\phi_{0}^{n}\left(\int_{\left(-\infty,0\right)}\sum_{l=1}^{L}d_{0,l}e^{-a_{0,l}s}dN^{n}\left(s\right)\right)\\
\leq & \left|g_{0}^{n}\right|_{\infty}\left(c_{0}+\mathbb{E}\int_{\left(-\infty,0\right)}\sum_{l=1}^{L}d_{0,l}e^{-a_{0,l}s}dN^{n}\left(s\right)\right)\\
= & \left|g_{0}^{n}\right|_{\infty}\left(c_{0}+\mathbb{E}\lambda^{n}\left(0\right)\int_{\left(0,\infty\right)}\sum_{l=1}^{L}d_{0,l}e^{-a_{0,l}s}ds\right).
\end{align*}
This is in the same form as the first display on page 1575 of BM96.
It is then easy to see that the above is finite if $\left|g_{0}^{n}\right|_{\infty}\int_{0}^{\infty}\sum_{l=1}^{L}d_{0,l}e^{-a_{0,l}s}ds<1$
which is the case if $\left|g_{0}^{n}\right|_{\infty}\sum_{l=1}^{L}\frac{d_{0,l}}{a_{0,l}}<1$.
Given that the intensity is not explosive for any $n$, ergodicity
of $N^{n}$ follows from the discussion on page 1573 of BM96. To show
that as $n\rightarrow\infty$ the intensity converges to $\lambda_{0}$
we use the same argument based on monotonicity, as done in the last
steps of the proof of Theorem 4 of BM96 on page 1575. This requires
that $\lambda^{n}$ and $N^{n}$ are increasing in $n$. Following
step by step the argument in BM96, this is true, here, if $g_{0}^{n}\left(t\right)$
is increasing in $n$. This is the case by the regularity conditions. 

We then use Theorem 1 in BM96 to see that the stationary distribution
is unique. For $x\geq0$, use $\phi_{t}\left(x\right)=g_{0}\left(t\right)\left(c_{0}+x\right)$
instead of fixed $\phi\left(x\right)$, as done there. We know that
$\phi_{t}\left(\cdot\right)$ is Lipschitz with Lipschitz constant
$\left|g_{0}\right|_{\infty}$ satisfying $\left|g_{0}\right|_{\infty}\int_{0}^{\infty}\sum_{l=1}^{L}d_{0,l}e^{-a_{0,l}s}ds<1$.
Then we can see that their proof follows through based on the remarks
above. This shows the stationarity of the process. 

The fact that the process restricted to $\mathcal{A}$ will converge
eventually to a stationary point process follows from Point b. in
Theorem 1 of BM96. The result holds for any process whose initial
condition satisfies $\sup_{t\geq0}\varepsilon_{v}\left(t\right)<\infty$
and $\lim_{t\rightarrow\infty}\varepsilon_{v}\left(t\right)=0$ almost
surely for all $v>0$, where $\varepsilon_{v}\left(t\right):=\int_{t-v}^{t}\int_{\left(-\infty,0\right)}\sum_{l=1}^{L}d_{0,l}e^{-a_{0,l}\left(r-s\right)}dN\left(s\right)dr$.
Clearly, this is the case for the initial condition $\mathcal{A}:=\left\{ N\left(t\right)=0:t\leq0\right\} $
that we are considering.

\subsection{Proof of Theorem \ref{Theorem_convergence}}

At first we state the following result on the moments of the process. 

\begin{lemma}\label{Lemma_moments}Under the Regularity Conditions,
for any finite interval $\left[r,s\right]$, $\mathbb{E}N^{p}\left(\left[r,s\right]\right)$
for any $p<\infty$, where $N\left(\left[r,s\right]\right):=\int_{r}^{s}dN\left(t\right)$.
Moreover, $\mathbb{E}\lambda_{0}^{4}\left(t\right)<\infty$ uniformly
in $t\geq0$.\end{lemma}

\begin{proof}This follows from a trivial modification of Lemmas 1
and 2 in Zhu (2013). Lemma 1 in Zhu (2013) says that nonlinear Hawkes
processes as in the proof of Theorem \ref{Theorem_stationarityErgodicity},
and that start at zero (i.e. conditioning on the empty past), have
moment of all order. By stationarity, Lemma 2 in Zhu (2013) extend
the result to the same Hawkes processes without conditioning on the
empty past. Following the proof of Lemma 1 in Zhu (2013), we can see
that the lemma also applies to our model as long as $\left|g_{0}\right|_{\infty}\leq B<\infty$.
In the proof of Lemma 1 in Zhu (2013), we just need to replace his
$\alpha$ and $\lambda\left(0\right)$ with our $B$ and $c_{0}B$,
respectively. Hence, our Hawkes process has moments of all orders. 

Finally, Lemma 15 in Guo and Zhu (2018) says that for a standard linear
Hawkes process the intensity has finite fourth moment. Following the
proof of Lemma 15 in Guo and Zhu (2018), using the fact that $\left|g_{0}\right|_{\infty}\leq B<\infty$,
we see that the result applies to our process as well. The proof requires
that $\mathbb{E}N^{4}\left(\left[r,s\right]\right)<\infty$ for some
suitably small interval $\left[r,s\right]$. This is the case by the
first statement of the present lemma.\end{proof}

Completing the square, we have that 
\begin{equation}
R_{T}\left(g;h\right)=\frac{1}{T}\int_{0}^{T}\left(\frac{\lambda_{0}}{h}-g\right)^{2}d\mu-\frac{1}{T}\int_{0}^{T}\left(\frac{\lambda_{0}}{h}\right)^{2}d\mu-\frac{2}{T}\int_{0}^{T}\frac{g}{h}dM.\label{EQ_RTRepresentation}
\end{equation}
Here, $dM=dN-\lambda_{0}d\mu$ so that $M$ is a martingale. Recall
that $\mathcal{G}=\left\{ g=X'b:\left|b\right|_{1}\leq B,b_{k}\geq0,k\leq K\right\} $.
By convexity, 
\begin{equation}
\sup_{b:b\geq0,\left|b\right|_{1}\leq B}\left|\frac{1}{T}\int_{0}^{T}\sum_{k=1}^{K}b_{k}X_{k}dM\right|\leq B\max_{k\leq K}\left|\frac{1}{T}\int_{0}^{T}X_{k}dM\right|.\label{EQ_L1Inequality}
\end{equation}
The main ingredient in the proof is to show convergence of $\max_{k\leq K}\sup_{h}\left|\frac{1}{T}\int_{0}^{T}\left(X_{k}/h\right)dM\right|$
to zero. This requires to control the oscillations of the process
over finite partitions of the parameter space of elements in $\mathcal{H}$.
To this end we use the following.

\begin{lemma}\label{Lemma_bracketingPartitionH} Let $h^{\left(j\right)}\left(t\right)=c^{\left(j\right)}+\sum_{l=1}^{L}h_{l}^{\left(j\right)}\left(t\right)$
where $h_{l}^{\left(j\right)}\left(t\right)=\int_{\left(-\infty,t\right)}d_{l}^{\left(j\right)}e^{-a_{l}^{\left(j\right)}\left(t-s\right)}dN\left(s\right)$,
$j=1,2$ such that $\left|c^{\left(1\right)}-c^{\left(2\right)}\right|\leq\epsilon$,
$\left|d_{l}^{\left(1\right)}-d_{l}^{\left(2\right)}\right|\leq\epsilon/\left(2L\right)$,
$\left|a_{l}^{\left(1\right)}-a_{l}^{\left(2\right)}\right|\leq\epsilon/\left(2\bar{d}L\right)$
where $a_{l}^{\left(1\right)},a_{l}^{\left(2\right)}\geq\underline{a}:=\min\left\{ a\in\mathcal{A}\right\} $
and $d_{l}d_{l}'\leq\bar{d}:=\max\left\{ d\in\mathcal{D}\right\} $.
Under the Regularity Conditions, for any arbitrary but fixed $h^{\left(1\right)}$
and $h^{\left(2\right)}$ satisfying the above, 
\begin{equation}
\frac{\int_{0}^{T}\left|h^{\left(1\right)}\left(t\right)-h^{\left(2\right)}\left(t\right)\right|^{2}\lambda_{0}\left(t\right)dt}{\epsilon^{2}}\lesssim T+\int_{0}^{T}\left[\int_{\left(-\infty,t\right)}e^{-a\left(t-s\right)}dN\left(s\right)\right]^{2}\lambda_{0}\left(t\right)dt\label{EQ_h1h2Bound}
\end{equation}
for any $a<\underline{a}$. For any $a\in\left(0,\underline{a}\right)$
we have that $\frac{1}{T}\int_{0}^{T}\left[\int_{\left(-\infty,t\right)}e^{-a\left(t-s\right)}dN\left(s\right)\right]^{2}\lambda_{0}\left(t\right)dt\rightarrow C_{a}$
in probability, where $C_{a}$ is a finite constant.\end{lemma}

\begin{proof}Write 
\begin{equation}
\left|h^{\left(1\right)}\left(t\right)-h^{\left(2\right)}\left(t\right)\right|\leq\left|c^{\left(1\right)}-c^{\left(2\right)}\right|+\sum_{l=1}^{L}\left|h_{l}^{\left(1\right)}\left(t\right)-h_{l}^{\left(2\right)}\left(t\right)\right|.\label{EQ_hMinusHPrimeDecomposition}
\end{equation}
By the mean value theorem and basic inequalities, 
\begin{align*}
\left|h_{l}^{\left(1\right)}\left(t\right)-h_{l}^{\left(2\right)}\left(t\right)\right|\leq & \int_{\left(-\infty,t\right)}\left|d_{l}^{\left(1\right)}-d_{l}^{\left(2\right)}\right|e^{-a_{l}^{\left(1\right)}\left(t-s\right)}dN\left(s\right)\\
 & +\left|a_{l}^{\left(1\right)}-a_{l}^{\left(2\right)}\right|\max_{\tau\in\left[0,1\right]}\int_{\left(-\infty,t\right)}d_{l}^{\left(2\right)}\exp\left\{ -a_{l,\tau}\left(t-s\right)\right\} \left(t-s\right)dN\left(s\right)
\end{align*}
where $a_{l,\tau}=a^{\left(2\right)}+\tau\left(a_{l}^{\left(1\right)}-a_{l}^{\left(1\right)}\right)$.
Now note that $e^{-x}$ is a decreasing function of $x\in\mathbb{R}$
and $e^{-x}x\leq Ce^{-\left(1-\epsilon\right)x}$ for any $\epsilon>0$
and some fixed $C<\infty$. Hence, For any $a<\underline{a}$, the
r.h.s. is bounded above by a constant multiple of 
\[
\left(\left|d_{l}^{\left(1\right)}-d_{l}^{\left(2\right)}\right|+\left|a_{l}^{\left(1\right)}-a_{l}^{\left(2\right)}\right|d_{l}^{\left(2\right)}\right)\int_{\left(-\infty,t\right)}e^{-a\left(t-s\right)}dN\left(s\right)\leq\frac{\epsilon}{L}\int_{\left(-\infty,t\right)}e^{-a\left(t-s\right)}dN\left(s\right).
\]
Inserting this bound in (\ref{EQ_hMinusHPrimeDecomposition}) gives
the first statement in the lemma. 

For any $r<t$, define the process $Y\left(r,t\right):=\left[\int_{\left(r,t\right)}e^{-a\left(t-s\right)}dN\left(s\right)\right]^{2}$
which is a measurable function of $N$. We claim that $\mathbb{E}\left|Y\left(-\infty,t\right)\right|^{2}$
and $\mathbb{E}\lambda_{0}^{2}\left(t\right)$ are finite. Moreover,
both quantities are stationary by Theorem \ref{Theorem_stationarityErgodicity}.
Then, Lemma 2 in Ogata (1978) says that $\frac{1}{T}\int_{0}^{T}Y\left(-\infty,t\right)\lambda_{0}\left(t\right)dt$
satisfies the ergodic theorem and converges in probability to a finite
constant, say $C_{a}$. This would prove the second statement in the
lemma. Hence, we need to show that the claims relating to the finite
moments are true. 

We show that $\mathbb{E}\left|Y\left(0,t\right)\right|^{2}<\infty$
and $\mathbb{E}\lambda_{0}^{2}\left(t\right)<\infty$, uniformly in
$t\geq0$. By stationarity of $N$ and monotonicity this implies that
$\sup_{t>0}\mathbb{E}\left|Y\left(-\infty,t\right)\right|^{2}<\infty$.
Select a constant $\Delta>0$ and partition the interval $\left(0,t\right]$
into $n\left(t\right)$ subintervals $\Delta_{i}:=\left(\left(i-1\right)\Delta,i\Delta\right]$
of size $\Delta$ with the $n\left(t\right)$ interval equal to $\Delta_{n\left(t\right)}:=\left(\left(n\left(t\right)-1\right)\Delta,t\right]$
. Define $Y_{i}=\left[\int_{\Delta_{i}}e^{-a\left(t-s\right)}dN\left(s\right)\right]^{2}$;
for ease of notation, in $Y_{i}$, we are suppressing the dependence
on $t$. Note that $Y_{i}\leq e^{-2a\left(t-\Delta i\right)}N^{2}\left(\Delta_{i}\right)$
if $i<n\left(t\right)$ and $Y_{n\left(t\right)}\leq N^{2}\left(\Delta_{n\left(t\right)}\right)$.
As in the main text, with some abuse of notation, $N\left(\Delta_{i}\right):=\int_{\Delta_{i}}dN\left(t\right)$.
Hence, $\mathbb{E}\left(Y\left(-\infty,t\right)\right)^{4}=\mathbb{E}\left(\sum_{i=1}^{n\left(t\right)}Y_{i}\right)^{4}\leq\left(1+\sum_{i=1}^{n\left(t\right)-1}e^{-2a\left(t-\Delta i\right)}\right)^{4}\mathbb{E}N^{4}\left(\Delta_{1}\right)$
using stationarity. By Lemma \ref{Lemma_moments}, $N\left(\Delta_{1}\right)$
has moments of all orders. Moreover, it is not difficult to see that
the sum in the parenthesis on the r.h.s. of the inequality is finite
uniformly in $t$. Hence, the first claim follows. Again, by Lemma
\ref{Lemma_moments}, $\mathbb{E}\lambda_{0}^{2}\left(t\right)<\infty$,
which is the last claim we needed to prove. Hence, the proof of the
present lemma is concluded.\end{proof}

We can then show the uniform convergence of the martingale process.

\begin{lemma}\label{Lemma_maximalXhIneqaulity}Under the Regularity
Conditions, 
\[
\mathbb{E}\sup_{h\in\mathcal{H}}\max_{k\leq K}\left|\int_{0}^{T}\frac{X_{k}}{h}dM\right|\lesssim\sqrt{T\ln\left(1+K\right)}.
\]
 \end{lemma}

\begin{proof} This is an application of Point (ii) in Theorem 2.5
in Nishiyama (2000). It is based on a number of conditions that we
shall verify. For each $\epsilon\in(0,\delta]$, with $\delta>0$
to be defined momentarily, we need to find a partition $\Pi\left(\epsilon\right)=\bigcup_{l=1}^{N\left(\epsilon,\Psi\right)}\Psi\left(\epsilon,l\right)$
of $\Psi=\mathcal{C}\cup\mathcal{A}\cup\mathcal{D}$ where $N\left(\epsilon,\Psi\right)$
is an integer valued function increasing in $\epsilon$, and such
that $N\left(\delta,\Psi\right)=1$; $N\left(\epsilon,\Psi\right)$
should not be confused with the point process $N$. We also need to
show that for constants $C_{1}$ and $C_{2}$ with probability going
to one, we have that
\begin{equation}
\int_{0}^{T}\max_{h\in\mathcal{H}}\max_{k\leq K}\left(\frac{X_{k}}{h}\right)^{2}\lambda_{0}d\mu\leq C_{1}T\label{EQ_predictableVariation}
\end{equation}
and 
\begin{equation}
\sup_{\epsilon\in(0,\delta]}\max_{1\leq l\leq N\left(\epsilon\right)}\int_{0}^{T}\max_{\psi,\phi\in\Psi\left(\epsilon,l\right)}\max_{k\leq K}\frac{\left|X_{k}\right|^{2}}{\epsilon^{2}}\left(\frac{1}{h_{\psi}}-\frac{1}{h_{\phi}}\right)^{2}\lambda_{0}d\mu\leq C_{2}T.\label{EQ_equicontinuity}
\end{equation}
At first, we show (\ref{EQ_predictableVariation}). To this end, note
that $\left|X_{k}\right|_{\infty}\leq1$, and that $h$ is bounded
below by $c_{\mathcal{H}}$. Then, the l.h.s. of (\ref{EQ_predictableVariation})
is bounded above by $\int_{0}^{T}c_{\mathcal{H}}^{-2}\lambda_{0}d\mu\leq2c_{\mathcal{H}}^{-2}P\lambda_{0}T$
with probability going to one; the factor $2$ can be reduced to any
arbitrary number greater than one. Hence, $C_{1}=2c_{\mathcal{H}}^{-2}P\lambda_{0}$. 

We now define the partition $\Pi\left(\epsilon\right)$ and use it
to verify (\ref{EQ_equicontinuity}). For each $\epsilon>0$ we define
a partition of the parameter space $\mathcal{C}=\left[\underline{c},\bar{c}\right]=\bigcup_{l=1}^{N\left(\epsilon,\mathcal{C}\right)}\mathcal{C}_{l}$,
$\mathcal{D}=\left[0,\bar{d}\right]^{L}=\bigcup_{l=1}^{N\left(\epsilon,\mathcal{D}\right)}\mathcal{D}_{l}$,
and $\mathcal{A}=\left[\underline{a},\bar{a}\right]^{L}=\bigcup_{l=1}^{N\left(\epsilon,\mathcal{A}\right)}\mathcal{A}_{l}$
such that $\left|c^{\left(1\right)}-c^{\left(2\right)}\right|\leq\epsilon$,
$\left\{ \left|d_{l}^{\left(1\right)}-d_{l}^{\left(2\right)}\right|\leq\frac{\epsilon}{2L}:l=1,2,...,L\right\} $,
$\left\{ \left|a_{l}^{\left(1\right)}-a_{l}^{\left(2\right)}\right|\leq\frac{\epsilon}{2\bar{d}L}:l=1,2,...,L\right\} $
for any parameters belonging to the same partition of $\mathcal{C}$,
$\mathcal{D}$ and $\mathcal{A}$, respectively. To avoid trivialities
in the notation, we are supposing that for each $l=1,2,...,L$, the
parameters are constrained in the same interval, e.g. $a_{l}\in\left[\underline{a},\bar{a}\right]$.
By the same steps used to establish (\ref{EQ_predictableVariation}),
the l.h.s. of (\ref{EQ_equicontinuity}) is bounded above by 
\[
\sup_{\epsilon\in(0,\delta]}\max_{1\leq l\leq N\left(\epsilon\right)}\int_{0}^{T}\max_{\psi,\phi\in\Psi\left(\epsilon,l\right)}\frac{1}{\epsilon^{2}}\left(\frac{h_{\psi}-h_{\phi}}{c_{\mathcal{H}}^{2}}\right)^{2}\lambda_{0}d\mu.
\]
By Lemma \ref{Lemma_bracketingPartitionH} we can choose $C_{2}\asymp c_{\mathcal{H}}^{-4}\left(1+C_{a}\right)$,
where $C_{a}$ is as defined there. The partition we use for $\mathcal{H}$
has cardinality $N\left(\epsilon,\mathcal{H}\right)=N\left(\epsilon,\mathcal{C}\right)N\left(\epsilon,\mathcal{D}\right)N\left(\epsilon,\mathcal{A}\right)\leq\left(\frac{\bar{c}-\underline{c}}{\epsilon}\right)\left(\frac{2L\bar{d}}{\epsilon}\right)^{L}\left(\frac{2L\left(\bar{a}-\underline{a}\right)}{\epsilon\bar{d}}\right)^{L}\leq\left(\delta/\epsilon\right)^{1+2L}$,
where $\delta=\max\left\{ \bar{c}-\underline{c},2L\bar{d},2L\left(\bar{a}-\underline{a}\right)/\bar{d}\right\} $.
This partition needs to be multiplied by $K$ because $\left\{ X_{k}:k=1,2,....,K\right\} $
is a family of processes with exactly $K$ elements. Hence, Theorem
2.5 Point (ii) in Nishiyama (2000) says that 
\[
\mathbb{E}\sup_{h\in\mathcal{H}}\max_{k\leq K}\left|\int_{0}^{T}\frac{X_{k}}{h}dM\right|\lesssim\sqrt{C_{2}T}\int_{0}^{\delta}\sqrt{\ln\left(\max\left\{ e,K\left(\frac{\delta}{\epsilon}\right)^{1+2L}\right\} \right)}d\epsilon+\frac{C_{1}T}{\delta\sqrt{C_{2}T}},
\]
where, for convenience, we slightly changed the definition of the
entropy function and we also simplified the statement of Theorem 2.5
Point (ii) in Nishiyama (2000) because (\ref{EQ_predictableVariation})
and (\ref{EQ_equicontinuity}) hold with probability going to one.
The integral is bounded above by a constant multiple of $\delta\sqrt{\ln\left(1+K\right)}$
and the result follows because $\delta$ is fixed.\end{proof}

We can now prove, one by one the points in the statement of Theorem
\ref{Theorem_convergence}. 

\paragraph{Proof of Point 1.}

This follows from (\ref{EQ_RTRepresentation}), (\ref{EQ_L1Inequality})
and Lemma \ref{Lemma_maximalXhIneqaulity} noting that $h\geq c_{\mathcal{H}}>0$. 

\paragraph{Proof of Point 2.}

The argument is standard (van der Vaart and Wellner, 2000, proof of
Theorem 3.2.5). For any $g,g'\in\mathcal{G}$ and $h\in\mathcal{H}$
define $\Delta\left(g,g^{h};h\right)=\frac{1}{T}\int_{0}^{T}\left(\frac{\lambda_{0}}{h}-g\right)^{2}d\mu-\frac{1}{T}\int_{0}^{T}\left(\frac{\lambda_{0}}{h}-g^{h}\right)^{2}d\mu$,
and $d^{2}\left(g,g'\right)=\frac{1}{T}\int_{0}^{T}\left(g-g'\right)^{2}d\mu$.
Note that $\Delta\left(g,g^{h};h\right)\geq\frac{1}{4}d^{2}\left(g,g^{h}\right)$
if $d^{2}\left(g,g^{h}\right)\geq4\frac{1}{T}\int_{0}^{T}\left(\frac{\lambda_{0}}{h}-g^{h}\right)^{2}d\mu$
(van der Vaart and Wellner, 2000, Problem 3.4.5). Clearly, if this
is not the case, $d^{2}\left(g,g^{h}\right)\leq4A_{T}$ and there
is nothing more to prove. Assuming that this is the case, given that
$\hat{g}^{h}$ minimizes $R_{T}\left(g;h\right)$, the event $d^{2}\left(\hat{g}^{h},g^{h}\right)>\epsilon$
is contained in the event 
\[
\sup_{g\in\mathcal{G}:d^{2}\left(g,g^{h}\right)>\epsilon}\left[R_{T}\left(g^{h};h\right)-R_{T}\left(g;h\right)\right]\geq0.
\]
Adding and subtracting $\Delta\left(g,g^{h};h\right)$ and using the
lower bound in terms of $d^{2}\left(g,g^{h}\right)$ deduce that the
event in the above display is contained in the event 
\[
\sup_{d^{2}\left(g,g^{h}\right)>\epsilon}\left|R_{T}\left(g^{h};h\right)-R_{T}\left(g;h\right)-\Delta\left(g,g^{h};h\right)\right|\geq\epsilon/4.
\]
By (\ref{EQ_RTRepresentation}), 
\[
R_{T}\left(g;h\right)-R_{T}\left(g^{h};h\right)-\Delta\left(g,g^{h};h\right)=-\frac{2}{T}\int_{0}^{T}\frac{g-g^{h}}{h}dM.
\]
From (\ref{EQ_L1Inequality}) and Lemma \ref{Lemma_maximalXhIneqaulity}
deduce that there is a finite constant $C$ such that the r.h.s. is
less than $C\times B\sqrt{T^{-1}\ln K}$ with probability going to
one. Hence, choosing $\epsilon=4C\times B\sqrt{T^{-1}\ln K}$ we deduce
that the probability of the event $\left\{ d^{2}\left(\hat{g}^{h},g^{h}\right)>\epsilon\right\} $
goes to zero when $\Delta\left(\hat{g}^{h},g^{h};h\right)\geq\frac{1}{4}d^{2}\left(\hat{g}^{h},g^{h}\right)$
and this concludes the proof of Point 2. 

\paragraph{Proof of Point 3.}

Using the previously defined notation and the fact that $g_{0}=\lambda_{0}/h_{0}$,
we have that $d^{2}\left(g^{\hat{h}},g_{0}\right)=\frac{1}{T}\int_{0}^{T}\left(\frac{\lambda_{0}}{h_{0}}-g^{\hat{h}}\right)^{2}$.
Adding and subtracting $\lambda_{0}/\hat{h}$ inside the square on
the r.h.s., and using a simple inequality, we have that
\begin{equation}
d^{2}\left(g^{\hat{h}},g_{0}\right)\leq\frac{2}{T}\int_{0}^{T}\left(\frac{\lambda_{0}}{\hat{h}}-g^{\hat{h}}\right)^{2}d\mu+\frac{2}{T}\int_{0}^{T}\left(\hat{h}-h_{0}\right)^{2}\left(\frac{\lambda_{0}}{\hat{h}h_{0}}\right)^{2}d\mu.\label{EQ_approxErrorBound}
\end{equation}
Similarly we deduce that 
\begin{equation}
\frac{1}{T}\int_{0}^{T}\left(\frac{\lambda_{0}}{\hat{h}}-g\right)^{2}d\mu\leq\frac{2}{T}\int_{0}^{T}\left(\frac{\lambda_{0}}{h_{0}}-g\right)^{2}d\mu+\frac{2}{T}\int_{0}^{T}\left(\hat{h}-h_{0}\right)^{2}\left(\frac{\lambda_{0}}{\hat{h}\hat{h}_{0}}\right)^{2}d\mu.\label{EQ_approxLossErrorBound}
\end{equation}
Taking $\inf_{g\in\mathcal{G}}$ on both sides and by definition of
$g^{\hat{h}}$, the above is equal to 
\begin{equation}
\frac{1}{T}\int_{0}^{T}\left(\frac{\lambda_{0}}{\hat{h}}-g^{\hat{h}}\right)^{2}\leq\frac{2}{T}\int_{0}^{T}\left(\hat{h}-h_{0}\right)^{2}\left(\frac{\lambda_{0}}{\hat{h}h_{0}}\right)^{2}d\mu,\label{EQ_approaxErrorUsingH}
\end{equation}
where we have used the fact that $g_{0}=\lambda_{0}/h_{0}\in\mathcal{G}$
and that the inf in the first term on the r.h.s. of (\ref{EQ_approxLossErrorBound})
is attained at $g_{0}$. Inserting the above display in (\ref{EQ_approxErrorBound})
and recalling that $\hat{h}\geq c_{\mathcal{H}}>0$,
\[
d^{2}\left(g^{\hat{h}},g_{0}\right)\leq\frac{4}{T}\int_{0}^{T}\left(\hat{h}-h_{0}\right)^{2}\left(\frac{g_{0}}{c_{\mathcal{H}}}\right)^{2}d\mu=O_{P}\left(B\sqrt{\frac{\ln\left(1+K\right)}{T}}\right),
\]
where the r.h.s. follows by assumption. Incidentally, using (\ref{EQ_approaxErrorUsingH})
we have also shown that $\frac{4}{T}\int_{0}^{T}\left(\frac{\lambda_{0}}{\hat{h}}-g^{\hat{h}}\right)^{2}d\mu=O_{P}\left(B\sqrt{\frac{\ln\left(1+K\right)}{T}}\right)$
so that we can just apply the result from Point 2 with $A_{T}=O_{P}\left(B\sqrt{\frac{\ln\left(1+K\right)}{T}}\right)$
and deduce Point 3 from the triangle inequality: $d\left(\hat{g}^{\hat{h}},g_{0}\right)\leq d\left(\hat{g}^{\hat{h}},g^{\hat{h}}\right)+d\left(g^{\hat{h}},g_{0}\right)$. 

\subsection{Proof of Theorem \ref{Theorem_QuadraticLossConsistency}}

For notational simplicity it is tacitly assumed that $g$ and $h$
are in $\mathcal{G}$ and $\mathcal{H}$. Subtracting $Q_{T}\left(h_{0},g_{0}\right)$
from $Q_{T}\left(h,g\right)$, 
\[
Q_{T}\left(h,g\right)-Q_{T}\left(h_{0},g_{0}\right)=\frac{1}{T}\int_{0}^{T}\left(\lambda_{0}-hg\right)^{2}d\mu-\frac{2}{T}\int_{0}^{T}hgdM,
\]
where $dM(t)=dN(t)-\lambda_{0}(t)dt$. By similar arguments as for
Lemma \ref{Lemma_maximalXhIneqaulity}, we have that 
\[
\mathbb{E}\sup_{h\in\mathcal{H}}\max_{k\leq K}\left|\int_{0}^{T}hX_{k}dM\right|\lesssim\sqrt{T\ln\left(1+K\right)}.
\]
This can be established noting that the analogue of (\ref{EQ_predictableVariation})
to the present case follows because $hX_{k}\lesssim\int_{\left(-\infty,t\right)}e^{-a\left(t-s\right)}dN\left(s\right)$
for any $a<\underline{a}$ as Lemma \ref{Lemma_bracketingPartitionH}.
The r.h.s. has finite expectation, as shown in the proof of Lemma
\ref{Lemma_bracketingPartitionH}. The analogue of (\ref{EQ_equicontinuity})
also follows using the arguments in the proof of Lemma \ref{Lemma_maximalXhIneqaulity}.
In consequence, the minimizers of $Q_{T}\left(h,g\right)$ w.r.t.
$h$ and $g$ minimize asymptotically $\frac{1}{T}\int_{0}^{T}\left(\lambda_{0}-hg\right)^{2}d\mu$.
By the Regularity Conditions, and similar arguments as in the proof
of Point 2 in Theorem \ref{Theorem_convergence}, we also deduce the
consistency rates of the estimator.

\subsection{Proof of Theorem \ref{Theorem_testCompetingIntensities}}

Using the definition of $\epsilon_{T}$ we have that 
\[
\frac{L_{T}^{\left(1\right)}-L_{T}^{\left(2\right)}}{\sqrt{T\hat{\sigma}_{T}^{2}}}=\frac{\int_{0}^{T}\left(\ln(\hat{\lambda}^{\left(1\right)})-\ln(\hat{\lambda}^{\left(2\right)})\right)dM}{\sqrt{T\hat{\sigma}_{T}^{2}}}+\frac{\epsilon_{T}}{\sqrt{T\hat{\sigma}_{T}^{2}}}
\]
where $dM(t)=dN(t)-\lambda_{0}(t)dt$. The second term on the r.h.s.
is $o_{p}\left(1\right)$ and can be disregarded. This is only true
if $\hat{\sigma}_{T}^{2}>0$ with probability going to one. By the
assumptions in the statement of the theorem, it is not difficult to
see that this is the case using ergodicity of $N$. Let $\Delta_{i}=\left(\left(i-1\right)P\lambda_{0},iP\lambda_{0}\right]$,
$i=1,2,...,n$ and to avoid trivialities suppose that $nP\lambda_{0}=T$.
We also note that $P\lambda_{0}=\left|\Delta_{i}\right|$, where the
r.h.s. is the Lebesgue measure of $\Delta_{i}$. Then, 
\begin{equation}
\frac{1}{\sqrt{T}}\int_{0}^{T}\left(\ln\left(\hat{\lambda}^{\left(1\right)}\right)-\ln\left(\hat{\lambda}^{\left(2\right)}\right)\right)dM=\frac{1}{\sqrt{n}}\sum_{i=1}^{n}Y_{i}\label{EQ_martingaleAverage}
\end{equation}
where $Y_{i}:=\left|\Delta_{i}\right|^{-1/2}\int_{\Delta_{i}}\ln\left(\frac{\hat{\lambda}^{\left(1\right)}}{\hat{\lambda}^{\left(2\right)}}\right)dM$.
By construction $\mathbb{E}_{i-1}Y_{i}=0$, where $\mathbb{E}_{i-1}$
is expectation conditioning on $\left(Y_{j}\right)_{j\le i-1}$. Given
that $\left(Y_{i}\right)_{i\geq1}$ is stationary and ergodic by Theorem
\ref{Theorem_stationarityErgodicity}, for the martingale central
limit theorem to apply to (\ref{EQ_martingaleAverage}), it is sufficient
that $\mathbb{E}\left|Y_{i}\right|^{2}<\infty$. Let $f=\ln\left(\frac{\hat{\lambda}^{\left(1\right)}}{\hat{\lambda}^{\left(2\right)}}\right)$
so that $\mathbb{E}\left|Y_{i}\right|^{2}=\left|\Delta_{i}\right|^{-1}\mathbb{E}\left(\int_{\Delta_{i}}fdM\right)^{2}$.
The r.h.s. is equal to $\left|\Delta_{i}\right|^{-1}\mathbb{E}\int_{\Delta_{i}}f^{2}\lambda_{0}d\mu$
by standard isometry. By stationarity, this is $Pf^{2}\lambda_{0}<\infty$.
To see this, use Holder inequality $Pf^{2}\lambda_{0}\leq Pf^{4}P\left|\lambda_{0}\right|^{2}$
and note that $Pf^{4}\lesssim P\hat{\lambda}^{\left(1\right)}+P\hat{\lambda}^{\left(2\right)}$
using a simple bound on the logarithm and the fact that the intensities
$\hat{\lambda}^{\left(k\right)}$ are bounded away from zero. Using
the same arguments as at the end of the proof of Lemma \ref{Lemma_bracketingPartitionH},
it is easy to deduce that $P\hat{\lambda}^{\left(k\right)}<\infty$.
This shows that (\ref{EQ_martingaleAverage}) converges to a normal
random variable with mean zero and variance $\mathbb{E}Y_{i}^{2}=\mathbb{E}f^{2}\lambda_{0}$.
Then, $\frac{1}{\sqrt{n}}\sum_{i=1}^{n}Y_{i}/\sqrt{\mathbb{E}f^{2}\lambda_{0}}$
has variance one, hence it is asymptotically standard normal in distribution.
By ergodicity of the counting process, $\mathbb{E}f^{2}\lambda_{0}=\hat{\sigma}_{T}^{2}+o_{p}\left(1\right)$
so that the assertion in (\ref{eq:T1}) is proved. 

\section{Empirical Study: Additional Details\label{Section_AppendixAdditionalDetails}}

We first discuss some of the nuisances and challenges of the dataset
we use. We then define the bins used for the construction of the one-hot
encoding of each variable. We conclude with details regarding the
parameters restrictions used in the estimation. 

\subsection{Data Conflation and Limits of Our Dataset\label{Section_dataConflationArrivals}}

Exchanges broadcast a number of information in the same packet via
the electronic communication network. This information is received
and consumed by any algorithm that subscribes to a given feed on the
network. The exchange timestamp does not allow to know what was sent
in a packet, as this is often a function of network capacity. Hence,
it is not possible to know how the information was received. Moreover,
the exchange timestamp and order id does not allow us to synchronize
data across multiple instruments with certainty. Historically, this
has been a clear problem with Level 2 data. However, we are not aware
of the extent to which such problem is mitigated by the use of Level
3 data on the NYSE. In depth knowledge usually requires analysis of
properly collected data in live trading. Here, we made the assumption
that all orders with the same exchange time stamp are sent and consumed
by an algorithm at the same time. We also synchronized different stocks
on the basis of their exchange timestamp. Finally, our data were queried
for 10 levels. It is not clear how this could be done in live trading.
In practice, the algorithm subscribing to a Level 3 feed is flooded
by all messages and conflation is necessary. This conflation may differ
substantially from what has been done in the present paper. 

No study that uses purchased data can escape the above problems. However,
to reassure the reader, we note that in a previous version of this
paper we carried out similar analysis using CME Level 2 data collected
by a proprietary trading firm colocated in the Aurora data centre
in Chicago. This data had nanosecond arrival timestamp (time at which
the data would have been seen by an algorithm) together with an identifier
for the order in which the information could be found within the same
packet. This allowed for perfect synchronization across products and
knowledge of the end of packet time stamp. It is reassuring that with
such dataset, we obtained results consistent with the  data used in
the present version of the paper. We do not report the results on
this high quality proprietary CME Level 2 data because it dates back
to 2014 and market microstructure might have changed in the meantime. 

\subsection{Parameters' Restrictions for Estimation\label{Section_paramsRestrictionsEmpirical}}

We impose the restriction that the linear coefficients $b_{k}$ are
in $\left[0,10\beta\right]$ where $\beta$ is as in Section \ref{EQ_BChoice}.
This implies that $B=10K\beta$. We multiply by 10 to avoid skrinking
coefficients too much. For the models that are linear in the raw covariates
we allow the linear coefficients $b_{k}$ to be in $\left[-10\beta,10\beta\right]$,
as otherwise we cannot capture negative impact. In this case, the
intensity is not guaranteed to be nonnegative. Hence, when testing,
we impose a lower bound on the intensity equal to ${\rm eps}$ for
the models linear in the raw covariates; ${\rm eps}$ is machine epsilon. 

Let $q_{{\rm Dur},10}$ and $q_{{\rm Dur},50}$ be the 10\% and 50\%
quantile of the trades durations. The parameters in the estimation
of $h$ in (\ref{EQ_hawkes_ht}) are restricted as follows, $c\in\left[10^{-3},10\right]/q_{{\rm Dur},50}$,
$d_{l}\in\left[0,1\right]/q_{{\rm Dur},50}$ and $a_{l}\in\left[10^{-2},10\right]/q_{{\rm Dur},10}$.
Scaling by $q_{{\rm Dur},10}$ and $q_{{\rm Dur},50}$ ensures that
we keep the correct order of magnitude irrespective of how time is
measured. Here, time is measured in seconds with nine decimal places
and for example $q_{{\rm Dur},10}$ tends to be in the order of $10^{-4}$. 
\end{document}